\documentclass[sigconf, natbib]{acmart}
\AtBeginDocument{%
  }

\setcopyright{acmlicensed}
\copyrightyear{2018}
\acmYear{2018}
\acmDOI{XXXXXXX.XXXXXXX}

\acmConference[Conference acronym 'XX]{Make sure to enter the correct
  conference title from your rights confirmation emai}{June 03--05,
  2018}{Ithaca, NY}
\acmISBN{978-1-4503-XXXX-X/18/06}

\usepackage{algorithm}
\usepackage{algorithmic}
\usepackage{xcolor}
\usepackage{pifont}
\usepackage[utf8]{inputenc}
\usepackage{natbib}
\usepackage{caption}
\usepackage{graphicx}
\usepackage{enumitem}
\usepackage{booktabs}
\usepackage{multirow}
\usepackage{amsmath} 
\usepackage{amsfonts}
\usepackage{amsthm}
\usepackage{bm}
\usepackage{tcolorbox}
\usepackage{subcaption}
\usepackage[normalem]{ulem}
\useunder{\uline}{\ul}{}

\newcommand{\nc}[1]{{\textbf{\color{black} [\textit{\color{green}\underline{\color{purple} NEED CITE!}}]}}}

\usepackage{newfloat}
\usepackage{listings}
\DeclareCaptionStyle{ruled}{labelfont=normalfont,labelsep=colon,strut=off} % DO NOT CHANGE THIS
\floatstyle{ruled}
\newfloat{listing}{tb}{lst}{}
\floatname{listing}{Listing}

\title{%CoReF: Communicative Reflectors with Fusion for Large Language Model-Based Sequential Recommendation
Decoding Recommendation Behaviors of In-Context Learning LLMs Through Gradient Descent
}

\newcounter{sharedfootnote}
\author{Yi Xu}
\authornote{The first two authors contributed equally to this research.}
\author{Weicong Qin}
\authornotemark[1]
\affiliation{%
  \department{Gaoling School of Artificial Intelligence}
  \institution{Renmin University of China}
  \city{}
  \country{}
}
\email{qwc@ruc.edu.cn}
\email{yixu00@ruc.edu.cn}

\author{Weijie Yu}
\authornote{Corresponding Author.}
\affiliation{%
\department{School of Information Technology and Management}
  \institution{University of International Business and Economics}
  \city{}
  \country{}}
\email{yu@uibe.edu.cn}

\author{Ming He}
\author{Jianping Fan}
\affiliation{%
  \institution{AI Lab at Lenovo Research}
  \city{}
  \country{}}
\email{heming01@foxmail.com}
\email{jfan1@lenovo.com}

\author{Jun Xu}
\affiliation{%
  \department{Gaoling School of Artificial Intelligence}
  \institution{Renmin University of China}
  \city{}
  \country{}
}
\email{junxu@ruc.edu.cn}

\renewcommand{\shortauthors}{Xu, Qin and Yu et al.}

\begin{abstract}

Recently, there has been a growing trend in utilizing large language models (LLMs) for recommender systems, referred to as LLMRec. A notable approach within this trend is not to fine-tune these models directly but instead to leverage In-Context Learning (ICL) methods tailored for LLMRec, denoted as LLM-ICL Rec. Many contemporary techniques focus on harnessing ICL content to enhance LLMRec performance.

However, optimizing LLMRec with ICL content presents unresolved challenges. Specifically, two key issues stand out: (1) the limited understanding of why using a few demonstrations without model fine-tuning can lead to better performance compared to zero-shot recommendations. (2) the lack of evaluation metrics for demonstrations in LLM-ICL Rec and the absence of the theoretical analysis and practical design for optimizing the generation of ICL content for recommendation contexts.

To address these two main issues, we propose a theoretical model, the \underline{L}LM-ICL \underline{R}ecommendation Equivalent \underline{G}radient \underline{D}escent model (LRGD) in this paper, which connects recommendation generation with gradient descent dynamics. We demonstrate that the ICL inference process in LLM aligns with the training procedure of its dual model, producing token predictions equivalent to the dual model's testing outputs. Building on these theoretical insights, we propose an evaluation metric for assessing demonstration quality. We integrate perturbations and regularizations in LRGD to enhance the robustness of the recommender system. To further improve demonstration effectiveness, prevent performance collapse, and ensure long-term adaptability, we also propose a two-stage optimization process in practice. Extensive experiments and detailed analysis on three Amazon datasets validate the theoretical equivalence and support the effectiveness of our theoretical analysis and practical module design. %Codes are available at \url{https://anonymous.4open.science/r/}.%\xy{Based on the auto-regressive output of decoder-only LLMs, we prove its equivalence with the output of the dual model through training-testing gradient descent. We also take factors like positional encoding, rotation matrices, and essential components into consideration, making the theory more complete. We integrate list-wise ranking metrics into the design of effective evaluation metrics for demonstrations, filling the gap between current ICL mechanism theory and LLM-ICL Rec. For practical applications, we xxx.} 

\end{abstract}



\begin{document}

\maketitle

% \showauthors@on

%     \link{Code}{https://anonymous.org/example/code}
%     \link{Datasets}{https://anonymous.org/example/datasets}
%     \link{Extended version}{https://anonymous.org/example/extended-version}

\section{Introduction}
Recently, large language models (LLMs) have emerged as promising recommenders due to their ability to capture complex user-item relationships from textual data, enabling them to generate contextually relevant recommendations. Unlike traditional recommendation models, which often rely on explicit user-item interactions or collaborative filtering signals, LLMs leverage their extensive pre-trained world knowledge and language understanding to offer more nuanced and dynamic recommendations. 

There are three primary approaches to exploring the recommendation potential of LLMs. The first, zero-shot learning (ZSL) methods~\cite{dai2023uncovering, hou2024large}, enable LLMs to generate recommendations without prior training or demonstrations, relying solely on their general knowledge. While this approach is computationally efficient, it struggles with personalization, failing to incorporate specific user behaviors or preferences. Fine-tuning methods~\cite{bao2023tallrec, liao2024llara, zheng2024adapting}, on the other hand, adapt the LLMs to recommendation tasks by training the model on a large labeled dataset. While this can improve performance, substantial computational resources are required for retraining. In contrast, few-shot learning (FSL) methods, also known as In-Context Learning based LLM recommendation (LLM-ICL Rec)~\cite{wang2023recmind, wang2024re2llm, qin2024enhancing} methods, leverage a few demonstrations to guide LLMs in making personalized recommendations without the need for retraining. This approach strikes a promising balance between recommendation performance and computational efficiency.

Despite the promising potential of ICL-based methods, several key challenges remain. \textbf{First}, there is a limited understanding of why a few demonstrations can significantly enhance performance without fine-tuning LLMs, leaving a gap in the theoretical grounding of how In-Context Learning influences LLM-based recommendations. \textbf{Second}, the lack of effective evaluation metrics for demonstrations hinders the ability to assess and compare the quality of different demonstration sets, making it difficult to determine which demonstrations lead to the best recommendations. \textbf{Third}, the absence of practical methods for optimizing demonstrations further complicates the process, as it becomes challenging to iteratively improve demonstrations for better recommendation outcomes. These challenges not only limit the efficacy of ICL-based recommender systems but also undermine their scalability and applicability.

% To address these challenges, we introduce a novel theoretical framework, the \underline{L}LM-ICL \underline{R}ecommendation Equivalent \underline{G}radient \underline{D}escent theoretical model (LRGD), for ICL-based recommendations, which connects recommendation generation with gradient descent dynamics. We demonstrate that the ICL inference process in LLM aligns with the training procedure of its dual model, producing token predictions equivalent to the dual model's testing outputs. 
% % QWC written
% Drawing upon these theoretical insights, we propose an evaluation metric for assessing demonstration quality. Furthermore, by incorporating  perturbation and regularization terms, we propose a two-stage iterative optimization process, which applies the LRGD to practice and thereby enhances the robustness of the recommender system. Extensive experiments and detailed analysis on three Amazon datasets confirm both the theoretical equivalence and the effectiveness of our analytical approach and practical design.
Facing the above challenges, we propose a novel model for ICL-based recommendations, which connects recommendation generation with gradient descent dynamics. We demonstrate that the generation of recommendation tokens in LLM-ICL Rec is mathematically equivalent to the gradient descent process of a dual model. Within this model, recommendation generation is treated as a training-testing process, where the LLM’s output is refined iteratively through gradient descent steps. This equivalence is extended from single-token to sequential token generation and generalized to both single-layer transformers and multi-layer decoder-only language models, showing that gradient descent dynamics hold consistently across different architectures. Unlike previous works~\cite{ren2024towards, dai2023why, zhang2023batch} that primarily analyze the mechanism of ICL under simplified settings, our model specifically focuses on recommendation settings and goes a step further by incorporating rotational positional encoding, sequential token generation processes, multi-layer transformer architectures, and multi-layer decoder-only language models. 

Additionally, building on this model, we introduce a novel evaluation metric, $\mathrm{Effect}_D$, to assess the quality of demonstrations systematically. This metric measures the efficiency of a demonstration by quantifying how quickly the model converges to the target item during the gradient descent process. 

To bridge theory with practice, we propose a two-stage iterative optimization process. In the first stage, the LLM generates new demonstrations based on user data. In the second stage, perturbation and regularization terms are applied to refine the demonstrations, enhancing their quality and robustness for practical recommendation scenarios.

Extensive experiments validate the LRGD model from both theoretical and practical perspectives. For the theoretical aspect, validation experiments confirm the equivalence and completeness of the LRGD theory. On the practical side, experiments on three real-world datasets demonstrate that LRGD achieves state-of-the-art performance compared to various recommendation methods, showcasing its practical effectiveness and model-agnostic adaptability.

% We demonstrate that the ICL inference process in LLM aligns with the training procedure of its dual model, producing token predictions equivalent to the dual model's testing outputs.

% \qwc{
% For the LRGD theory, we conduct validation experiments to confirm the equivalence and the completeness. For the LRGD application, we conduct extensive experiments on three real-world datasets. LRGD achieves state-of-the-art performance against various recommendation methods, demonstrating the practical effectiveness and model-agnostic superiority of LRGD. Additionally, Ablation experiments confirm the effectiveness of additional designs in LRGD.}

In summary, the contributions of this paper are as follows:
\begin{itemize}[leftmargin=*]
    \item We introduce the LRGD model, which for the first time establishes the equivalence between recommendation token generation in LLM-ICL Rec and gradient descent dynamics of a dual model, providing theoretical insights into why demonstrations enhance performance without fine-tuning. 
    % \item \qwc{Unlike previous theoretical works, LRGD for the first time bridges the gap between theoretical foundations and practical applications of LLM-ICL Rec. Specifically, we innovatively take key components of decoder-only models (e.g., positional encoding, rotation matrices, etc.) into consideration and extend LRGD to sequential token generation, single-layer transformers, and multi-layer decoder-only models, which demonstrates its generality and alignment with the LLM backbone.}
    \item We propose a novel evaluation metric, $\mathrm{Effect}_D$, to systematically assess demonstration quality by measuring its impact on gradient descent convergence, bridging theoretical understanding with practical demonstration optimization.
    \item We design a two-stage iterative optimization process integrating perturbation and regularization terms to ensure the robust and scalable LLM-ICL Rec for real-world recommendation applications. 
    % \item Theoretical experiment result confirms the equivalence of our theoretical analysis on LRGD. Extensive experiment results on three real-world datasets demonstrate the state-of-the-art performance of LRGD. Detailed analysis of each additional design's effectiveness shows the reliability and model-agnostic superiority of LRGD.
\end{itemize}

\section{Related Works}

\subsection{LLM-Based Recommendation}
Traditional recommendation models often rely on DNN-based modules, including convolutional and recurrent neural networks~\cite{hidasi2015session,tang2018personalized,li2017neural} and multi-layer perceptron-based models~\cite{li2022mlp4rec,li2023automlp,zhou2022filter}. Transformer-based models~\cite{kang2018self,sun2019bert4rec,he2021locker,he2022query} further improve item-to-item relevance through self-attention, setting new benchmarks.

Despite their success, these methods struggle to capture the full complexity and fluidity of user preferences due to limited world knowledge and reasoning ability. As a result, LLM-based methods have emerged as a promising solution. Current LLMRec approaches fall into two categories: fine-tuning and prompt-based methods. Fine-tuning methods~\cite{geng2022recommendation} treat recommendation as a question-answering task and fine-tune accordingly, incorporating recommendation knowledge through special tokens (e.g., LLaRA~\cite{liao2024llara}, LC-Rec~\cite{zheng2024adapting}) or text embeddings (e.g., A-LLMRec~\cite{kim2024large}). However, fine-tuning can be costly and inefficient for practical use~\cite{qin2024enhancing}. 

Prompt-based methods, such as ZSL Rec and LLM-ICL Rec, offer alternatives. For instance, LLM4RS~\cite{dai2023uncovering} customizes prompts to enhance LLM’s recommendation capabilities, while LLMRank~\cite{hou2024large} leverages zero-shot ranking through specialized prompts. Recent works like \citet{wang2024re2llm} and \citet{qin2024enhancing} use reflection-based approaches to enhance future recommendations by leveraging LLMs to analyze interaction history. Despite their efforts, these prompt-based methods often lack clear explanation for their prompt design, as well as theoretical analysis.

\subsection{Understanding the Mechanism of ICL}

% \xy{Recently, several works have attempted to understand the inference mechanism of in-context learning (ICL) in decoder-only models. Brown et al.~\cite{brown2020language} shows the remarkable ability of ICL. Many works try to understand ICL via empirical methods like simple text functions or Bayesian inference~\cite{garg2022can, xie2022explanation,  wang2024large}. Many other works~\cite{schlag2021linear, akyurek2022learning, bai2024transformers} also link ICL with gradient descent, aiming to explore transformer’s ability to perform gradient descent algorithms to achieve ICL. Dai et al.~\cite{dai2023why} and Zhang et al.~\cite{zhang2023batch} utilize the dual form to understand ICL as a gradient descent of the original model under a linear attention setting. Ren et al.~\cite{ren2024towards} extend linear attention setting to softmax via kernel approach. 

% However, as Shen et al.~\cite{shen2024position} pointed out that, current methods do not consider the impact of positional encoding and rotation matrix, or omitting some important parts like multi-layer transformers, perturbations and regularizations. Also, there is still a big gap for these mechanism to be applied in RecSys.}
Recently, several works have attempted to understand the inference mechanism of in-context learning (ICL) in decoder-only models. \citet{brown2020language} show the remarkable ability of ICL. Many studies~\cite{garg2022can, xie2022explanation, wang2024large} have explored ICL through empirical methods, such as simple text functions or Bayesian inference. Other works~\cite{schlag2021linear, akyurek2022learning, bai2024transformers} connect ICL with gradient descent, investigating the transformer’s capacity to execute gradient descent algorithms for ICL. \citet{dai2023why} and \citet{zhang2023batch} use a dual form to understand ICL as a gradient descent of the original model under a linear attention framework. \citet{ren2024towards} extend the linear attention model to softmax through a kernel approach.

However, as \citet{shen2024position} point out, current methods overlook factors like positional encoding, rotation matrices, and essential components such as multi-layer transformers, perturbations, and regularizations. Additionally, there remains a significant gap in applying these mechanisms to recommender systems.
% \input{2_preliminary}
% \section{Methodology}
\section{LRGD: The Proposed Model}
We propose the LRGD model and practical strategies used for LLM-ICL Rec in this section.
% In this section, we propose LRGD, to better understand the theoretical mechanism of LLM-ICL Rec (Sec. ~\ref{sec:method:theory}-\ref{sec:method:extention}). We also define the effectiveness of demonstrations in LLM-ICL Rec with analysis and explanations (§\ref{sec:method:theory:effective}), introduce additional designs for LRGD (§\ref{sec:method:extra}) , and give practical recommendation's application based on LRGD and deal problems we faced in practical recommendation scenarios (§\ref{sec:method:app}).
\subsection{Problem Formulation and Notations}
\label{sec:method:problem}
\begin{figure}[!t]
    \setlength{\belowcaptionskip}{0pt}
    \setlength{\abovecaptionskip}{2pt}
    \centering
    \includegraphics[width=0.4\textwidth]{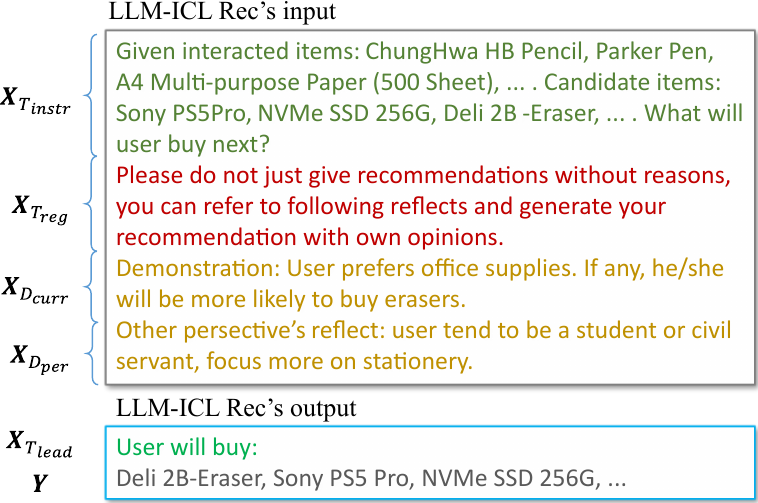}
    \caption{Input and output segment example. Text with different colors corresponding to different part of $\bm X$ and $\bm Y$.}
    \label{fig:input:example}
\end{figure}

In this study, we focus on the LLM-ICL Rec problem. Formally, given an input sequence \(\bm X\) containing user-specific information, the objective is to generate a ranked list of recommended items, denoted as \(\bm Y\). The recommendation is formulated as follows:
\[
\bm Y = \text{LLM-ICL}_{\text{rec}}(\bm X).
\]
As illustrated in Fig.~\ref{fig:input:example}, the input \(\bm X\) consists of two main components: \(\bm X = [\bm X_T, \bm X_D]\). The task instruction \(\bm X_T\) includes the basic instructions for the recommendation task (\(\bm X_{T_{\text{instr}}}\)) and the sequence of tokens generated prior to the current recommendation (\(\bm X_{T_{\text{lead}}}\)), which may include reasoning or explanations. The demonstration \(\bm X_D\) contains the current demonstration (\(\bm X_{D_{\text{curr}}}\)), reflecting the user's preferences or relevant information from past interactions.
\begin{figure}[!t]
    \setlength{\belowcaptionskip}{0pt}
    \setlength{\abovecaptionskip}{2pt}
    \centering
\includegraphics[width=0.45\textwidth]{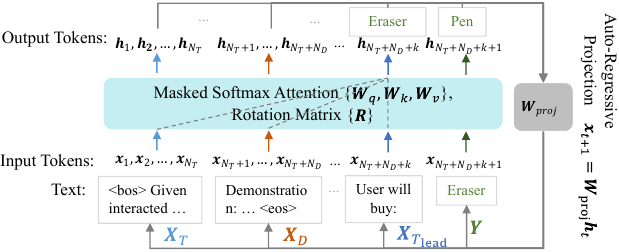}
    \caption{Basic inference mechanism of LLM-ICL Rec}
    \label{fig:basic}
\end{figure}

The recommendation is achieved through an auto-regressive generation process. For example, consider the generation of the first recommended item in \( \bm{Y} \) (i.e. `Eraser' in Fig~\ref{fig:input:example}), corresponding to the \( k \)-th generated output token \( \bm h_{t(k)} \), where \( t(k) = N_T+N_D+k\) denotes its index. 
This generation is informed by the preceding tokens, including \( \bm{X_D} \) (with length \( N_D \)), \( \bm{X}_{T_{\text{lead}}} \) (with length \( k \)), and the \( \bm{X_T} \)  excluding \( \bm{X}_{T_{\text{lead}}} \) (with length \( N_T \)).  These tokens are processed through the attention mechanism, where the query, key, and value matrices are denoted as \( \bm W_i \in \mathbb{R}^{d_o \times d_i} \), for \( i \in \{q, k, v\} \). In this case, the token \( \bm h_{t(k)} \) is generated using the last input token from \( \bm X_{T_{\text{lead}}} \) as the query vector \( \bm q \) and \( t(k) - 1 \) preceding tokens, i.e., $[\bm x_{i}]_{i=1}^{t(k)-1}$ as the value and key vectors $\bm V, \bm K$. For simplicity, we define the index sets as $\mathcal{I} = \mathcal{I}_T \cup \mathcal{I}_D$, where $\mathcal{I}_T = [i]_{i=1}^{N_T} \cup [i]_{i=t(0)}^{t(k-1)}$ represents the indices for $\bm{X_T}$, and $\mathcal{I}_D = [i]_{i=N_T+1}^{N_T+N_D}$ represents the indices for $\bm{X_D}$, These index sets allow for the clear identification of the tokens contributing to the value and key matrices in the attention mechanism.

Moreover, unlike existing works~\cite{dai2023why, ren2024towards} that overlook positional encoding, we incorporate it into our theoretical analysis to better align with the practical applications of LLM-ICL Rec. Specifically, we inject the positional encoding into \( \bm W_q \) and \( \bm W_k \) using Rotation Positional Encoding (RoPE)~\cite{su2024roFormer}. This is achieved by multiplying a rotation matrix \( \bm R_i \in \mathbb{R}^{d_o \times d_o} \), which satisfies \( \bm R_m^\top \bm R_n = \bm R_{m-n}, \forall m, n \in \mathbb{N_+} \). The value matrix \( \bm W_v \) is left unchanged. The current input token is then transformed into the query vector as \( \bm q = \bm R_{t(k)} \bm W_q \bm x_{t(k)} \in \mathbb{R}^{d_o} \).%, where \( t(k) = N_T + N_D + k \) denotes the token index. 

In the following subsections, we
(1) theoretically explore ICL-based recommendation from the perspective of gradient descent (§\ref{sec:method:problem:analysis}-§\ref{sec:method:theory}).
(2) propose an evaluation metric to assess the effectiveness of demonstrations for ICL-based recommendations (§\ref{sec:method:theory:effective}).
(3) extend \(\bm X_T\) and \(\bm X_D\) with regularization (\(\bm X_{T_{\text{reg}}}\)) and perturbation (\(\bm X_{D_{\text{per}}}\)), respectively, to improve recommendation performance (§\ref{sec:method:extra}).
(4) provide practical solutions for applying our theoretical analysis to real-world recommendation scenarios (§\ref{sec:method:app}).
\subsection{Connect Attention with Gradient Descent}
\label{sec:method:problem:analysis}
\subsubsection{\textbf{Connection between Attention and Gradient Descent}}

Recent studies \cite{ren2024towards, dai2023why, zhang2023batch, yao2024enhancing} establish duality between gradient descent on linear layers and linear attention. The single-layer linear model is defined as:
\begin{equation}
f(x) = \bm W_0 \bm x
\end{equation}
Given train input sequence $\bm X=[\bm x_k]_{k=1}^N$ and labels $[\bm y_k]_{k=1}^N$, weights are updated via error signals $\bm E =[\bm e_k]_{k=1}^N = [-\beta \partial\mathcal{L}/\partial\bm y_k]_{k=1}^N$:
\begin{equation}
\bm W' = \bm W_0 + \sum_{k=1}^{N} \bm e_k \otimes \bm x_k
\end{equation}
Linear attention ($\mathrm{LA}$) can be expressed as:
\begin{equation}
\mathrm{LA}(\bm V,\bm K,\bm q) = \left( \sum_{k=1}^{N} \bm v_k \otimes \bm k_k \right)\bm q
\end{equation}
For test input $\bm x'$, the equivalent form combines gradient descent and linear attention:
\begin{equation}
f(\bm x') = \bm W' \bm x'= \bm W_0 \bm x' + \mathrm{LA}(\bm E,\bm X,\bm x')
\end{equation}

Meanwhile, following \cite{ren2024towards}, softmax attention is approximated via kernel methods to match linear attention. 
% Detailed description and observation can be found in Part A of \url{https://anonymous.4open.science/r//auxiliary.pdf}.
Detailed description and observation can be found in Sec~\ref{app:conn}.

For individual vectors $\bm x, \bm y \in \mathbb{R}^{d_o}$, we can rewrite $\mathrm{exp}(\bm x, \bm y)$ with random Fourier mapping $\phi(\bm \cdot)$ based on the approximation of common RBF kernel~\cite{rahimi2007random} as:
\begin{equation}
\label{eq:pre:exp}
\mathrm{exp}(\bm x, \bm y) = e^{\bm x^\top \bm y} = \phi(\bm x)^\top \phi(\bm y),
\end{equation}
% where 
% $$
% \phi(\bm x) = \frac{ e^{||\bm x||^2_2 / 2}}{\sqrt{D}} \left( \sin(\bm u_1^\top \bm x), .., \sin(\bm u_{D/2}^\top \bm x), \cos(\bm u_1^\top \bm x), .., \cos(\bm u_{D/2}^\top \bm x) \right)^\top,
% $$
% and \(D\) is a constant (typically around 100), with \(\bm u_i\) being random vectors drawn from \(\mathcal{N}(0, \sigma\bm I_{d_o})\). The mapping \(\phi(\bm x)\) transforms \(\bm x \in \mathbb{R}^{d_o}\) to \(\mathbb{R}^{d_D}\). $\phi(\bm \cdot)$ is consistent with the form proposed in \cite{choromanski2020rethinking} and will be utilized in this paper.
% After applying the $\phi(\bm \cdot)$ to the individual vectors in \(\bm X_1\) and \(\bm X_2\) as described above, we obtain:
% $$
% \exp(\bm X_1^\top \bm X_2) = \phi(\bm X_1)^\top \phi(\bm X_2).
% $$
Thus, the approximate form of the softmax function is:
\begin{equation}
    \mathrm{softmax}(\bm x^\top \bm y) = c\exp(\bm x^\top \bm y) = c\phi(\bm x)^\top \phi(\bm y) ,
\end{equation}
where \(c = (\phi(\bm x)^\top \phi(\bm y))^{-1} \in \mathbb{R}^{1}\).

Finally, the output of the softmax attention can be rewritten as:
\begin{equation}
\label{eq:pre:kernel}
    \bm h = \bm V \, \mathrm{softmax}\left(\frac{\bm K^\top \bm q}{\sqrt{d_o}}\right) = c\bm V \, \phi(\bm K)^\top \phi(\bm q) = \mathrm{LA}(c \bm V, \phi(\bm K)^\top, \phi(\bm q)),
\end{equation}
where
$c = \left( \bm 1_{N}^\top \phi(\bm K)^\top \phi(\bm q) \right)^{-1} \in \mathbb{R}^1
$, $\bm q \in \mathbb{R}^{d_i}$ is query vector.

This demonstrates how softmax attention in transformers can be approximated using kernel methods, transforming the softmax component into a linear attention form.

\subsection{Gradient Descent in LLM-ICL Recommend}
\label{sec:method:theory}
We illustrate the basic inference mechanism of LLM-ICL Rec in Fig.~\ref{fig:basic} and introduce LRGD.

\subsubsection{\textbf{Understanding LLM-ICL Recommendation with Gradient Descent}}
\label{sec:method:theory:understand}
% We follow \cite{ren2024towards} and present the output of LRGD in the context of a simplified single-layer masked softmax attention and demonstrate its equivalence to gradient descent theory. 
% In LLM-ICL Rec, as illustrated in Fig.~\ref{fig:basic}, the input token \( \bm x_{t+1} \) at time step \( t+1 \) is auto-regressively generated from the hidden state \( \bm h_t \) at time step \( t \). This relationship is captured by a projection matrix \( \bm W_{\text{proj}} \in \mathbb{R}^{d_i \times d_o} \), i.e., \( \bm x_{t+1} = \bm W_{\text{proj}} \bm h_t \).
We consider the generation of  \( \bm h_{t(k)} \in \mathbb{R}^{d_o} \), the first output item in \( \bm Y \). 
As illustrated in Fig.~\ref{fig:basic}, this generation involves the attention mechanism, which is closely associated with gradient descent dynamics in its dual form, as discussed in Eq.~\eqref{eq:pre:kernel} in §\ref{sec:method:problem:analysis}. The computation is expressed as:
\begin{equation}
\label{eq:method:basic:softmax}
\bm h_{t(k)} = \bm V \, \mathrm{softmax}\left(\frac{\bm K^\top \bm q}{\sqrt{d_o}}\right) = c \bm V \phi(\bm K)^\top \phi(\bm q),
\end{equation}
where $d_o$ denotes the output token dimension, $\phi(\cdot)$ denotes the softmax kernel, and the components are defined as follows:
\begin{itemize}[leftmargin=*]
\item
\( c = (\bm 1_{N_T+N_D}^\top \phi(\bm K)^\top \phi(\bm q))^{-1}\),
\item
\( \bm V = [\bm V_T, \bm V_D] = [\bm W_v \bm X_T, \bm W_v \bm X_D] = [[\bm v_i]_{i \in \mathcal{I}_T}, [\bm v_j]_{j \in \mathcal{I}_D}] \), 
\item
\( \bm K = [\bm K_T, \bm K_D] = [[\bm R_i \bm W_k \bm x_i], [\bm R_j \bm W_k \bm x_j]] = [[\bm k_i]_{i \in \mathcal{I}_T}, [\bm k_j]_{j \in \mathcal{I}_D}] \).
\end{itemize}
The computation can be further decomposed into the contributions from task instructions \( \bm X_T \) and demonstrations \( \bm X_D \):
\begin{equation}
    \bm h_{t(k)} = c \bm V_T \phi(\bm K_T)^\top \phi(\bm q) + c \bm V_D \phi(\bm K_D)^\top \phi(\bm q).
\end{equation}
The dual model associated with $\bm h_{t(k)}$ can be proved as:
\begin{equation}
\label{eq:dual-ht}
  f(\bm q) = \bm W \phi(\bm q) = \bm W_0 \phi(\bm q) - \mathrm{grad} \cdot \phi(\bm q), 
\end{equation}
where \( \bm W_0 = c \bm V_T \phi(\bm K_T)^\top \) is a constant matrix dependent only on \( \bm X_T \), and \( \mathrm{grad} \) is the gradient associated with the demonstrations \( \bm X_D \). The corresponding \( \mathrm{grad} \) and loss function \( \mathcal{L}_{ICL} \) are given as:
\begin{equation}
\label{eq:dual-grad}
    \mathrm{grad} = -c \bm V_D \phi(\bm K_D)^\top,\\
    \mathcal{L}_{ICL} = -\frac{c}{\beta} \sum_{i \in \mathcal{I}_D} (\bm v_i)^\top (\bm W \phi(\bm k_i)),
\end{equation}
where \( \beta \) is the effective learning rate for the dual model. 
% Detailed proof can be found in Part B of \url{https://anonymous.4open.science/r//auxiliary.pdf}.
Detailed proof can be found in Sec~\ref{app:proof}.

%%%%%%%%%%%%%%%%%%%%
% \textsc{Proof.~}\textit{To derive the gradient of \( \mathcal{L}_{ICL} \) with respect to \( \bm W \), we first compute the derivative:} 
% \begin{equation}
% \frac{\partial \mathcal{L}_{ICL}}{\partial \bm W} = -\frac{c \partial \sum_{i \in \mathcal{I}_D} (\bm v_i)^\top \bm W \phi(\bm k_i)}{\beta \partial \bm W} = -\frac{c}{\beta} \sum_{i \in \mathcal{I}_D} \bm v_i \phi(\bm k_i)^\top.
% \end{equation}

% \textit{Thus, the gradient \( \mathrm{grad}_{ICL} \) is equivalent to $\mathrm{grad}$:}
% \begin{equation}
% \mathrm{grad}_{ICL} = \beta \frac{\partial \mathcal{L}_{ICL}}{\partial \bm W} = -c \bm V_D \phi(\bm K_D)^\top = \mathrm{grad}.
% \end{equation}

% \textit{Consequently, the output of the dual model aligns with $\bm h_{t(k)}$:
% }
% \begin{equation}
% \label{eq:equal}
% \begin{aligned}
% f(\bm q) &= \bm W \phi(\bm q) = {\bm W}_0 \phi(\bm q) - \mathrm{grad} \cdot \phi(\bm q) \\&= c \bm V_T \phi(\bm K_T)^\top \phi(\bm q) + c \bm V_D \phi(\bm K_D)^\top \phi(\bm q) = \bm h_{t(k)}.  
% \end{aligned}
% \end{equation}
%%%%%%%%%%%%%%%%%%%%%
This confirms that the generation process of $\bm h_{t(k)}$ in LRGD is equivalent to the dual model's gradient descent. 

\begin{figure}[!t]
    \centering
    \setlength{\belowcaptionskip}{0pt}
    \setlength{\abovecaptionskip}{2pt}
    \includegraphics[width=0.45\textwidth]{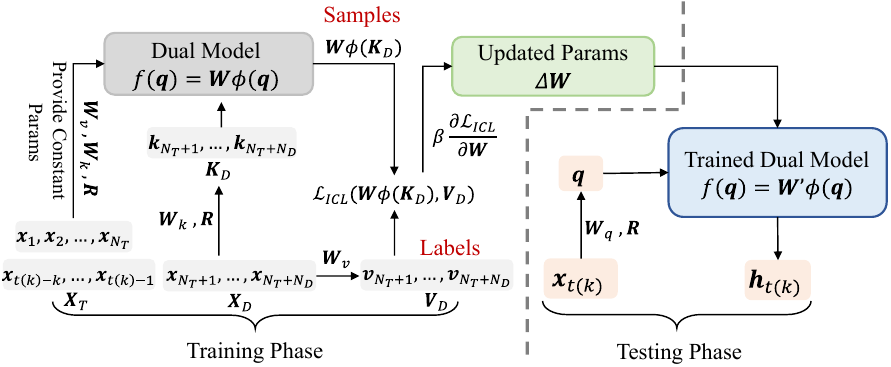}
    \caption{Gradient descent mechanism in dual model}
    \label{fig:equity}
\end{figure}

Inspired by~\cite{ren2024towards, dai2023why, zhang2023batch}, the process for LRGD generating tokens can be viewed as a training-testing round in the dual model, as shown in Fig.~\ref{fig:equity}:

(1) \textbf{Training Phase}: Fix \( \bm X_T \) and use \( \bm X_D \) to construct label-sample pairs. Here, the labels are \( \bm V_D = [\bm v_i]_{i \in \mathcal{I}_D} \), and the samples are \( \bm W \phi(\bm K_D) = [\bm W \phi(\bm k_i)]_{i \in \mathcal{I}_D} \). The loss function computes the cosine similarity between labels and samples. After computing \( \mathcal{L}_{ICL} \) and its gradients, a single step of stochastic gradient descent is performed, updating \( \bm W \) to \( \bm W' \).

(2) \textbf{Testing Phase}: Use the updated dual model to generate the next token based on the new query \( \bm q \), resulting in the output \( \bm h_{t(k)} \).

This process confirms the equivalence between LRGD token generation and gradient descent dynamics.  Further theoretical verification is conducted in §~\ref{sec:exp:theo}.

\subsubsection{\textbf{Extending Equivalence to Sequential Token Generation}}
The analysis above can be naturally extended to the sequential generation of multiple tokens. Specifically, 
as shown in Eq.~\eqref{eq:dual-ht} and Eq.~\eqref{eq:dual-grad}, the dual model's gradient descent is influenced by \( \bm X_T \) and \( \bm X_D \), in other words, the generated outputs from previous steps in \( \bm Y \) are treated as constants when generating the next token. 
For each new token generation, the corresponding new round of training-testing is updated through 
$\bm W_0'=\bm W_0 + c\bm v_{t(k+1)}\phi(\bm k_{t(k+1)})$ and
$\phi(\bm q')=\phi(\bm R_{t(k+1)}\bm W_q\bm x_{t(k+1)})$ as detailed in Eq.~\eqref{eq:dual-ht}, thereby shifting the starting point for the gradient descent. Gradient descent is implemented by leveraging sample-label pairs illustrated in Fig.~\ref{fig:equity}. These pairs are obtained from $\bm X_D$ and are used during the generation of each new output token by the dual model.

% Each new token generation corresponds to a new round of training-testing, where the constant matrix \( \bm W_0 \) and the input variable \( \phi(\bm q) \) are updated. These updates are driven by the introduction of a new input token \( \bm x_{t+1} = \bm W_{\text{proj}} \bm h_t \) as defined by the auto-regressive mechanism, thereby shifting the starting point for the next gradient descent iteration. During training, the demonstration \( \bm X_D \) provides label-sample pairs, enabling the gradient descent.

In summary, the sequential token generation process seamlessly aligns with the gradient descent equivalence established earlier. It systematically extends the single-token mechanism to multi-token outputs while also providing a foundation for designing metrics to evaluate the quality of ICL demonstrations, as detailed in §\ref{sec:method:theory:effective}.

% The above conclusion can be extended to generating multiple new tokens sequentially: as defined, the newly generated outputs in former steps are treated as constants before generating the next token. Each new token generation involves a new round of training-testing, where the constant matrix \( \bm W_0 \) and the input variable \( \phi(\bm q) \) are updated due to the new input \( \bm x_{t+1} = \bm W_{\text{proj}} \bm h_t \) introduced by the auto-regressive mechanism, changing the starting point for the next gradient descent. The training process uses the previous \( \bm X_D \), along with the corresponding loss function and gradients, to perform one step of gradient descent. In essence, for each new token in LLM-ICL Rec, \( \bm X_D \) serves as the provider of training label-sample pairs for the next token's gradient descent. If the target is not reached, the dual model continues performing gradient descent based on \( \bm X_D \) until the target is achieved or the gradient descent steps reach the limit. This also informs the design for measuring the quality of ICL demonstrations in §\ref{sec:method:theory:effective}.

% \subsection{Extention to Language Models}
% \label{sec:method:extention}
% We further generalize LRGD to the following scenarios: single-layer transformer, and multi-layer decoder-only language model to align with real-world applications of LLM-ICL Rec~\cite{bao2023tallrec,liao2024llara,qin2024enhancing}. The corresponding dual models, gradients, and loss function equivalences are provided for each case in the following two parts.
%%%%%%%%%%%%%%%%%%%%%%
\subsubsection{\textbf{Generalization to Single-Layer Transformer and Multi-Layer Decoder-Only Language Models}}
\label{sec:method:trm}
%The dual model corresponding to a single-layer transformer is given by $ f(\bm q) = \bm W_{\text{trm}}(\bm q) + \bm b_{\text{trm}} $, and the equivalent gradient descent and loss function for the next input token can be derived from the following process. 
We further generalize LRGD to single-layer transformer to facilitate the analysis of multi-layer decoder-only language models. Based on Eq.~\eqref{eq:method:basic:softmax} and output form of transformer, we analyze the generation of the $ k $-th output token $\hat{\bm x}_{t(k)}$ for a single-layer transformer as follows:
\begin{equation}
\begin{aligned}
&\hat{\bm x}_{t(k)} = \bm W_{\text{FFN}_1} \left[ \bm \Sigma_{\text{act}} \left( \bm W_{\text{FFN}_2} \bm h_{t(k)} + \bm b_{\text{FFN}_2} \right) \right] + \bm b_{\text{FFN}_1},
%&= \bm W_{\text{trm}} \bm V_T \phi(\bm K_T)^\top \phi(\bm q) + \bm W_{\text{trm}} \bm V_D \phi(\bm K_D)^\top \phi(\bm q) + \bm W_{\text{FFN}_1} \bm \Sigma_{\text{act}} \bm b_{\text{FFN}_2} + \bm b_{\text{FFN}_1}\\
%&= \bm W_{\text{trm}}' \phi(\bm q) - \beta_{\text{trm}} \frac{\partial \mathcal{L}}{\partial \bm W_{\text{trm}}} \phi(\bm q) + \bm b_{\text{trm}}
\end{aligned}
\end{equation}
% where $ \bm W_{\text{trm}} = c_{\text{trm}} \bm W_{\text{FFN}_1} \bm \Sigma_{\text{act}} \bm W_{\text{FFN}_2} $,
% $ \bm W_{\text{trm}}' = \bm W_{\text{trm}} \bm V_T \phi(\bm K_T)^\top $,
% $ \bm b_{\text{trm}} = \bm W_{\text{FFN}_1} \bm \Sigma_{\text{act}} \bm b_{\text{FFN}_2} + \bm b_{\text{FFN}_1} $,
% $ c_{\text{trm}} = \left( \bm 1_{N_T+N_D}^\top \phi(\bm K)^\top \phi(\bm q) \right)^{-1} $.
where $\bm W_{\text{FFN}_1} \in \mathbb{R}^{d_o \times d_h}$, $\bm b_{\text{FFN}_1} \in \mathbb{R}^{d_o}$, $\bm W_{\text{FFN}_2} \in \mathbb{R}^{d_h \times d_o}$, $\bm b_{\text{FFN}_2} \in \mathbb{R}^{d_h}$ are the transformer FFN parameters, and $\bm \Sigma_{\text{act}} \in \mathbb{R}^{d_h \times d_h} $ is the
equivalent mapping of activation functions\footnote{The activation function can be ReLU, SwiGLU, etc. 
Following~\cite{ren2024towards}, we treat it as a constant once $ \bm X_D $ and $ \bm X_T $ are fixed.}.

The corresponding dual model for a single-layer Transformer  can then be expressed as: 
\begin{equation}
\label{eq:dual-st}
f(\bm q) = \bm W_{\text{trm}}\phi(\bm q) + \bm b_{\text{trm}} = \bm W_{\text{trm},0}\phi(\bm q)-\mathrm{grad}_{\text{trm}}\phi(\bm q)+\bm b_{\text{trm}}.
\end{equation}
The constants of $f(\bm q)$ are $\bm W_{\text{trm},0} = \hat{\bm W}_{\text{trm}} \bm V_T \phi(\bm K_T)^\top $, $ \bm b_{\text{trm}} =\bm b_{\text{FFN}_1} + \bm W_{\text{FFN}_1} \bm \Sigma_{\text{act}} \bm b_{\text{FFN}_2}$, where $\hat{\bm W}_{\text{trm}} = c_{\text{trm}} \bm W_{\text{FFN}_1} \bm \Sigma_{\text{act}} \bm W_{\text{FFN}_2} $. Thus, the corresponding gradient $\mathrm{grad}_{\text{trm}}$ and loss function $\mathcal{L}_{\text{trm}}$ are:
\begin{equation}
\begin{aligned}
\mathrm{grad}_{\text{trm}} &= -\hat{\bm W}_{\text{trm}} \bm V_D \phi(\bm K_D)^\top = - \sum_{i \in \mathcal{I}_D} \left( \hat{\bm W}_{\text{trm}} \bm v_i \right) \otimes \phi(\bm k_i), \\
\mathcal{L}_{\text{trm}} &= -\frac{1}{\beta_{\text{trm}}} \sum_{i \in \mathcal{I}_D} \left( \hat{\bm W}_{\text{trm}} \bm v_i \right)^\top \left( \bm W \phi(\bm k_i) + \bm b \right).
\end{aligned}
\end{equation}
Single-layer transformer's generation process of $\hat{\bm x}_{t(k)}$ is equivalent to the gradient descent of dual model introduced in Eq.~\eqref{eq:dual-st} through prove procedure similar with Eq.~\eqref{eq:dual-ht}. 

Furthermore, to align with real-world applications of LLM-ICL Rec~\cite{bao2023tallrec,liao2024llara,qin2024enhancing}, we generalize LRGD to multi-layer decoder-only language models with  $L$  layers (GPT\cite{achiam2023gpt},  LLaMa\cite{dubey2024llama}, etc.). 
% Detailed observation of the generalization of LRGD for multi-layer decoder-only language models can be found in Part C \& D of~\url{https://anonymous.4open.science/r//auxiliary.pdf}.
Detailed observation of the generalization of LRGD for multi-layer decoder-only language models can be found in Sec~\ref{app:obs:multi:layer}
 and Sec~\ref{app:obs:GQA}.

\subsection{Evaluation of Effective ICL Demonstrations in Recommendation}
\label{sec:method:theory:effective}
% As analyzed in §\ref{sec:method:theory:understand}, we have proven that the sequential token generation in LLM-ICL Rec is equivalent to multiple gradient descent operations. This theoretical analysis reveals that high-quality demonstrations ($\bm X_D$) play a critical role in guiding the gradient updates and improving the alignment of model predictions with the target outputs. In this context, the position of the ground truth item in the output token list directly reflects the effectiveness of the demonstrations. Intuitively, when the ground truth item appears earlier in the output list, it indicates that the model has better aligned its gradient updates to the user's preferences, as encoded in the demonstrations.
As discussed in §\ref{sec:method:theory:understand}, the sequential token generation in LLM-ICL Rec is equivalent to multiple gradient descent operations. High-quality demonstrations ($\bm X_D$) are crucial for guiding gradient updates and aligning model predictions with the target outputs. To evaluate demonstration quality, we propose the metric $\mathrm{Effect}_{D}$:
\begin{equation}
\mathrm{Effect}_{D} = \frac{1}{\log_2(i+1)},
\end{equation}
where $i$ is the position of the ground truth item token in the LLM’s output token list. This metric penalizes later positions and reflects the efficiency of the gradient descent process in reaching the target.

\begin{figure}[!t]
    \centering
    \setlength{\belowcaptionskip}{0pt}
    \setlength{\abovecaptionskip}{5pt}
    \includegraphics[width=0.48\textwidth]{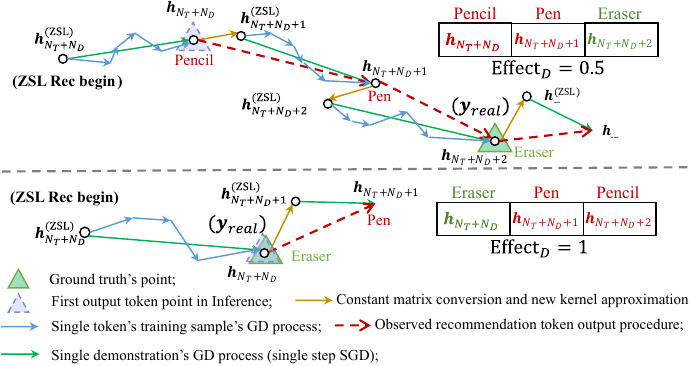}
    \caption{Visualization of the evaluation metric $\mathrm{Effect}_D$.}
    \label{fig:eval:metric}
\end{figure}

Fig.~\ref{fig:eval:metric} illustrates the effectiveness of $\mathrm{Effect}_{D}$ in measuring demonstration quality. 
The space represents the label space of the dual model \( f(\bm q) = \bm W \phi(\bm q) \), with distance measured by mean squared error. $\bm X_{T_{\text{lead}}}$ is omitted for simplicity. 
In the upper part of Fig.~\ref{fig:eval:metric}, when $\mathrm{Effect}_{D}=0.5$
the ground truth item ``Eraser'' appears as the third token, requiring three gradient descent steps to reach. In contrast, in the lower part, $\mathrm{Effect}_{D}=1$ the ground truth appears as the first token, reducing the number of steps to one. This demonstrates that better demonstrations accelerate convergence, aligning with our theoretical analysis.
Moreover, in practical machine learning scenarios where the validation and test distributions are similar, a higher $\mathrm{Effect}_{D}$ on the validation set implies better generalization performance on the test set. Therefore, using \( \mathrm{Effect}_{D} \) to evaluate demonstration quality helps identify better demonstrations, providing useful metrics and guidance for our design in §\ref{sec:method:app}.

We conduct theoretical verification experiments in §\ref{sec:exp:theo} to validate the practical effectiveness of the proposed metric.

% \subsubsection{Preliminary validation experiment for LIRE-GDTM}
% \label{sec:method:theory:understand}

% We conduct a preliminary validation experiment for our proposed LIRE-GDTM to validate its correctness...

% However, there are still some issues with finding excellent demonstrations based on \( \mathrm{Effect}_{D} \): (1) The assumption that the distribution difference between the validation and test sets is small is difficult to meet in practical applications. (2) When two demonstrations result in the same \( \mathrm{Effect}_{D} \) value, it is hard to further distinguish which one has better practical recommendation performance (e.g., better generalization ability). In the next section, we will propose additional designs to address these two issues.

% We conduct theoretical verification experiments in §\ref{sec:exp:theo} to validate the practical effectiveness of the metrics.

\subsection{Extensional Terms for LRGD}
In practical applications, leveraging $\mathrm{Effect}_{D}$ to find optimal demonstrations poses two challenges:
\begin{itemize}[leftmargin=*]
\item The assumption that the validation and test set distributions are similar may not hold in real-world scenarios.
\item When two demonstrations yield the same $\mathrm{Effect}_{D}$ scores, it becomes difficult to determine which one has superior practical recommendation performance, such as better generalization.
\end{itemize}
To address these issues, we propose Perturbations \( \bm X_{D_{per}} \) and Regularizations  \( \bm X_{T_{reg}} \).
We analyze how these designs affect the gradient descent dynamics of the dual model, enabling more robust recommendation performance in §\ref{sec:method:app}.

% \xy{However, there are still some issues with finding excellent demonstrations based on \( \mathrm{Effect}_{D} \) in §\ref{sec:method:theory:effective}: (1) The assumption that the distribution difference between the validation and test sets is small is difficult to meet in practical applications. (2) When two demonstrations result in the same \( \mathrm{Effect}_{D} \) value, it is hard to further distinguish which one has better practical recommendation performance (e.g., better generalization ability). %In the next section, we will propose additional designs to address these two issues.

% To address issues mentioned above, we provide additional designs for \( \bm X_T \) and \( \bm X_D \) of the LRGD model:  (1) Perturbations for \( \bm X_D \); (2) Regularizations for \( \bm X_T \). We analyze the impact of both on the gradient descent process of the dual model and the corresponding recommendation process, facilitating the proposal of the actual application of LRGD in §\ref{sec:method:app}. }

% Therefore, we can introduce actual application of LRGD in §\ref{sec:method:app} through these additional designs.

\label{sec:method:extra}
\subsubsection{\textbf{Perturbations}}
\label{sec:method:extra:perturb}
\begin{figure}[!t]
    \centering
    \setlength{\belowcaptionskip}{0pt}
    \setlength{\abovecaptionskip}{0pt}
    \includegraphics[width=0.45\textwidth]{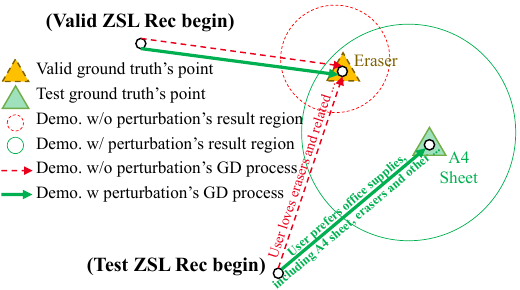}
    \caption{Mechanism of perturbation. `Demo.' represents for `demonstration'. `w/o' represents for `without'. `w/' represents for `with'. `GD' represents for `gradient descent'.}
    \label{fig:perturbation}
\end{figure}

Incorporating \( \bm X_{D_{per}} \) into \( \bm X_{D} \) influence the dual model's gradient descent process. For example, as shown in Fig.~\ref{fig:input:example}, additional ICL demonstrations are introduced as perturbations. These perturbations act as supplements or corrections, adding an extra loss term to the objective function:
\begin{equation}
    \mathcal{L}_{ICL} = -\frac{c}{\beta} \left[\sum_{i \in \mathcal{I}_D} \bm v_i^\top (\bm W \phi(\bm k_i)) + \sum_{j \in \mathcal{I}_{D_{per}}} \bm v_j^\top (\bm W \phi(\bm k_j))\right].
\end{equation}
The effect of the perturbation, shown in Fig.~\ref{fig:perturbation}, is similar to dropout or random noise in the dual model. It alters the gradient descent direction, guiding it towards a broader result region that includes the ground-truth result. This enhances the generalization ability of LRGD, preventing it from converging to a narrow result region limited to the validation set. Instead, it enables the model to adapt to potential shifts in the ground-truth result region on the test set, improving its generalization to new test samples.

\subsubsection{\textbf{Regularizations}}
\label{sec:method:extra:regularize}

We propose two regularization methods to further enhance LLM-ICL Rec.

First, following \cite{ren2024towards}, we regularize the value head of \( \bm X_T \) in  the masked softmax attention by introducing a coefficient $(1-\alpha)$:
\begin{equation}
    h_{t(k)}' = [(1-\alpha)\bm V_T, \bm V_D] \mathrm{softmax}\left(\frac{\bm K^\top \bm q}{\sqrt{d_o}}\right),
\end{equation}
which corresponds to controlling the magnitude of gradient descent by introducing a regularization parameter into the loss function in the dual model:
\begin{equation}
   \mathcal{L}'_{ICL} = -\frac{c}{\beta} \sum_{i \in \mathcal{I}_D} (\bm v_i)^\top (\bm W \phi(\bm k_i)) + \frac{\alpha}{2\beta} ||\bm W||_F^2. 
\end{equation}
The added L2 regularization effectively reduces model variance, leading to improved generalization and robust recommendations.

Second, considering that manipulating the value head during inference is difficult to implement and inconsistent with the LLM pre-training in practice~\cite{ren2024towards, shi2024medadapter}, we propose a simpler method by directly introducing a regularization term
\( \bm X_{T_{\text{reg}}} \) to $\bm X_{T}$, as illustrated in Fig.~\ref{fig:input:example}.
The term $\bm X_{T}$ modifies the initial point of the dual model before gradient descent, providing a better starting position. This adjustment helps avoid poor local minima and ensures convergence toward the correct recommendation target. However, \( \bm X_{T_{\text{reg}}} \) may lead to significant fluctuations in recommendation outcomes. Therefore, regularization should not be frequently altered or directly modified.
\begin{figure}[!t]
    \centering
    \setlength{\belowcaptionskip}{0pt}
    \setlength{\abovecaptionskip}{5pt}
    \includegraphics[width=0.44\textwidth]{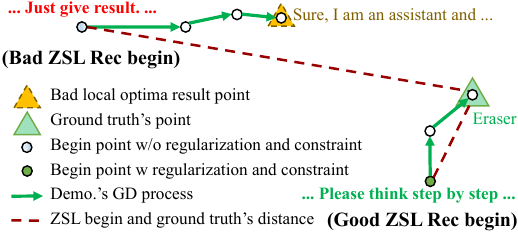}
    \caption{Mechanism of regularization}
    \label{fig:regularization}
\end{figure}

\subsection{Applying LRGD to LLM-ICL Rec}
\label{sec:method:app}
Given $\mathrm{Effect}_D$ in §\ref{sec:method:theory:effective} and extensional terms in Sec.\ref{sec:method:extra}, this section details how to apply LRGD and address issues in practical recommendation scenarios. 
\begin{figure}[!t]
    \centering
    \setlength{\belowcaptionskip}{0pt}
    \setlength{\abovecaptionskip}{5pt}
    \includegraphics[width=0.45\textwidth]{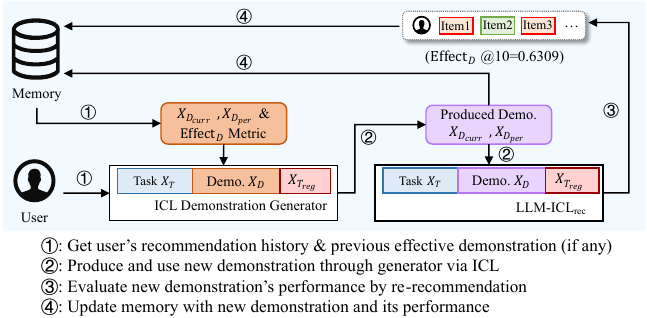}
    \caption{Detailed two-stage iterative optimization workflow for LRGD in practical recommendation scenarios.}
    \label{fig:2:phase:rec}
\end{figure}
As depicted in Fig.~\ref{fig:2:phase:rec}, applying our proposed LRGD model to practical recommendation involves a two-stage optimization process:
\begin{enumerate}[leftmargin=*]
    \item ICL Demonstration Generator (stage-1): The LLM-based generator constructs new demonstrations by leveraging user behaviors, profiles, and previously effective demonstrations (if any) stored in memory.
    \item LLM-ICL Rec (stage-2): The LLM recommender produces recommended items using the demonstrations constructed in stage 1. Simultaneously, the proposed evaluation metric $\mathrm{Effect}_D$ is employed to assess the quality of the demonstration. If the demonstration leads to correct recommendations, it is added to the memory for future iterations.
\end{enumerate}
The aim of the demonstration generator is to generate better demonstrations to further improve $\mathrm{Effect}_D$ at the most recent timestamp. Previous effective demonstrations are evaluated and selected through the corresponding $\mathrm{Effect}_D$ at that time.

However, iteratively optimizing demonstrations in this manner often leads to the ``collapse'' phenomenon~\cite{shumailov2024ai}, where the improvement margin from optimized demonstrations diminishes, or the optimized demonstrations become increasingly similar to one another. We will investigate this phenomenon in the experiments.
Since LLM-ICL Rec is connected to gradient descent dynamics as shown in §\ref{sec:method:theory:understand}, we argue that the cause of this ``collapse'' lies in the  ``local optima'' and ``error accumulation" problem\cite{dang2021escaping, parthipan2024defining}, i.e., iteratively optimized demonstrations gradually accumulate inaccuracies, resulting in deviations from the actual user preferences.

To address this issue, we integrate the proposed $X_{D_{\text{per}}}$ and $X_{T_{\text{reg}}}$ in §\ref{sec:method:extra} into the iterative optimization process to mitigate collapse and enhance robustness. Specifically, we define a reasonable $\bm X_{D_{\text{per}}}$ for a demonstration should be: (1) aimed at the same user; (2) 
generated independently of the original demonstration's optimization. Thus, we set up $m$-path demonstration optimization ($m$-PDO), where each path optimizes demonstrations independently.
When we detect the ``collapse'' phenomenon in the iteration, we randomly choose a demonstration from another path as $\bm X_{D_{\text{per}}}$ to provide new insights and prevent ``collapse''. Furthermore, we designate a good $\bm X_{T_{\text{reg}}}$ as a textual constraint that limits the output token range of the LLM.

\section{Experiment}
In this section, we conduct extensive experiments to answer the following research questions: 
\textbf{RQ1:} How equivalent are the decoder-only model and its dual model with gradient descent regarding output?
% How does the equivalence between the output of the decoder-only model and the gradient descent of the dual model in LRGD work? 
\textbf{RQ2:} How effective is the LRGD application compared to existing methods? 
\textbf{RQ3:} Can LRGD enhance the recommendation performance of various decoder-only LLMs? 
\textbf{RQ4:} How effective are the extensional terms in LRGD?
% Are the modules designed based on LRGD, such as demonstration rewriting iterations, perturbations, and regularization terms effective? 
\textbf{RQ5:} Can $\bm X_{D_{per}}$ in LRGD alleviate the demonstration collapse issue (§\ref{sec:method:theory:effective})?
% \xy{Fish king please help!}
% {sec:method:theory:effective}

% \begin{itemize}
%     \item \textbf{RQ1:} How does the equivalence between the output of the decoder-only model and the gradient descent of the dual model in LRGD work?
%     \item \textbf{RQ2:} How effective is the LLM-ICL Rec method based on LRGD in practice?
%     \item \textbf{RQ3:} Are the modules designed based on LRGD, such as demonstration rewriting iterations, perturbations, and regularization terms effective?
%     \item \textbf{RQ4:} Has the demonstration collapse problem encountered in practice been alleviated by using LRGD?
% \end{itemize} 

\subsection{Experiment Settings}
\textbf{Dataset.} LRGD is applicable to various LLM recommendation scenarios, e.g. general and sequential recommendation. For convenience, we choose sequential recommendation to validate the effectiveness of LRGD. Following ~\cite{zheng2024adapting}, we conduct our experiments on three subsets of the Amazon Review dataset (2018)~\cite{ni2019justifying}: ``Arts'', ``Video'', and ``Instruments''\footnote{``Arts'' refers to ``Arts, Crafts and Sewing''. ``Games'' refers to ``Video Games''. ``Instruments'' refers to ``Musical Instruments''.}. We follow the preprocessing strategy in~\cite{zheng2024adapting,qin2024enhancing}, including 5-core user-item interaction filtering, with the detailed statistics of the pre-processed datasets presented in Tab.~\ref{tab:dataset}.\\
\textbf{Baselines.}
For the baselines, we considered both traditional and large language model based approaches.

\begin{table}[!tbp]
\centering
\setlength{\belowcaptionskip}{-2pt}
\setlength{\abovecaptionskip}{5pt}
\caption{Statistics of the 3 pre-processed datasets.}
\label{tab:dataset}
{\small\begin{tabular}{lcccc}
\hline
Dataset       & \#Users   & \#Items     & Sparsity \\ \hline
Arts    & 55970           & 22612                         & 99.96\%          \\
Games   & 55145           & 17287                         & 99.95\%          \\
Instruments & 27404        & 10450                         & 99.92\%          \\ \hline
\end{tabular}}
\end{table}
\begin{table*}[tbp!]
\centering
\caption{Overall comparison of recommendation performance. LLM-based models all use LLaMa-3 as the backbone. The best and second-best results are displayed in bold and underlined fonts, respectively. ``Imp'' indicates the percentage improvement of LRGD over the best baseline performance. ``*'' denotes the improvement of LRGD is statistically significant (\(p \leq 0.05\)).}
\label{tab:exp:overall}
{\small\begin{tabular}{@{}c|c|ccccc|ccccc|c@{}}
\hline
                              &                          & \multicolumn{5}{c|}{Traditional Recommendation Models}                                     & \multicolumn{5}{c|}{LLM-Based Recommendation Models}       &                                   \\
\cline{3-12}
\multirow{-2}{*}{Dataset}     & \multirow{-2}{*}{Metric} & HGN    & SRGNN  & GRU4Rec & FDSA                          & SASRec                        & LC-Rec & E4SRec & Re2LLM & MoRE         & LRGD            & \multirow{-2}{*}{Imp.} \\ \hline
                              & nDCG@1                   & 0.1107 & 0.0939 & 0.0941  & {\color[HTML]{262626} 0.1195} & {\color[HTML]{262626} 0.1985} & 0.2566 & 0.2283 & 0.2193 & {\ul 0.2633} & \textbf{0.2977} & 13.06\%*                          \\
                              % & nDCG@5                   & 0.2231 & 0.1875 & 0.2007  & 0.2179                        & 0.3047                        & 0.3499 & 0.3557 & 0.3238 & {\ul 0.3681} & \textbf{0.3997} & 8.58\%*                           \\
                              & nDCG@10                  & 0.2587 & 0.2416 & 0.2627  & 0.2544                        & 0.3002                        & 0.3698 & 0.3720 & 0.3537 & {\ul 0.4045} & \textbf{0.4306} & 6.45\%*                           \\
                              % & Recall@5                 & 0.3291 & 0.2826 & 0.3089  & 0.2994                        & 0.3684                        & 0.4229 & 0.4109 & 0.4201 & {\ul 0.4689} & \textbf{0.4975} & 6.10\%*                           \\
\multirow{-3}{*}{Games}       & Recall@10                & 0.4407 & 0.4496 & 0.4459  & 0.4576                        & 0.4562                        & 0.4560 & 0.4699 & 0.5164 & {\ul 0.5601} & \textbf{0.6004} & 7.20\%*                           \\ \hline
                              & nDCG@1                   & 0.1077 & 0.1077 & 0.0994  & 0.1249                        & {\color[HTML]{262626} 0.1833} & 0.2126 & 0.2123 & 0.2676 & {\ul 0.2942} & \textbf{0.3005} & 2.14\%*                           \\
                              % & nDCG@5                   & 0.1812 & 0.1763 & 0.1662  & 0.1792                        & {\color[HTML]{262626} 0.2917} & 0.3034 & 0.3101 & 0.3396 & {\ul 0.3705} & \textbf{0.3756} & 1.38\%*                           \\
                              & nDCG@10                  & 0.2131 & 0.2182 & 0.2090  & 0.2263                        & {\color[HTML]{262626} 0.2909} & 0.3223 & 0.3384 & 0.3670 & {\ul 0.3922} & \textbf{0.3962} & 1.75\%*                           \\
                              % & Recall@5                 & 0.2526 & 0.2431 & 0.2353  & 0.2645                        & {\color[HTML]{262626} 0.3404} & 0.3689 & 0.3855 & 0.4075 & {\ul 0.4316} & \textbf{0.4420} & 2.41\%*                           \\
\multirow{-3}{*}{Arts}        & Recall@10                & 0.3508 & 0.3737 & 0.3796  & 0.3767                        & {\color[HTML]{262626} 0.4291} & 0.4406 & 0.4716 & 0.4865 & {\ul 0.4984} & \textbf{0.5063} & 1.59\%*                           \\ \hline
                              & nDCG@1                   & 0.1355 & 0.1358 & 0.1283  & 0.1426                        & {\color[HTML]{262626} 0.1877} & 0.1598 & 0.1603 & 0.1967 & {\ul 0.2213} & \textbf{0.2496} & 12.79\%*                          \\
                              % & nDCG@5                   & 0.2180 & 0.2197 & 0.2300  & 0.2064                        & {\color[HTML]{262626} 0.2548} & 0.2469 & 0.2674 & 0.2781 & {\ul 0.2918} & \textbf{0.3213} & 10.11\%*                          \\
                              & nDCG@10                  & 0.2522 & 0.2523 & 0.2636  & 0.2488                        & {\color[HTML]{262626} 0.2839} & 0.2727 & 0.2980 & 0.3010 & {\ul 0.3252} & \textbf{0.3500} & 7.63\%*                           \\
                              % & Recall@5                 & 0.4021 & 0.2976 & 0.2873  & 0.2777                        & {\color[HTML]{262626} 0.3201} & 0.3267 & 0.3705 & 0.3477 & {\ul 0.3717} & \textbf{0.3918} & 5.41\%*                           \\
\multirow{-3}{*}{Instruments} & Recall@10                & 0.2940 & 0.3986 & 0.3894  & 0.3752                        & {\color[HTML]{262626} 0.3729} & 0.3838 & 0.4428 & 0.4293 & {\ul 0.4686} & \textbf{0.4776} & 1.92\%*                           \\ \hline
\end{tabular}}
\end{table*}
(1) HGN~\cite{ma2019hierarchical} captures both short \& long-term user interests via hierarchical gating networks. 
(2) SRGNN~\cite{wu2019session} models session-based recommendations using graph neural networks for transitions.
(3) GRU4Rec~\cite{hidasi2015session} utilizes Gated Recurrent Units (GRUs) to model sequential user behavior for the session-based recommendation.
(4) FDSA~\cite{zhang2019feature} employs feature-level self-attention to capture dynamic patterns in user behavior.
(5) SASRec~\cite{kang2018self} leverages self-attention mechanisms to capture user preferences in sequential data.
(6) LC-Rec~\cite{zheng2024adapting} fine-tunes an LLM for recommendation with a variety of alignment tasks. 
(7) E4SRec~\cite{li2023e4srec} integrates LLMs with ID-based recommendations for efficiency.
(8) Re2LLM~\cite{wang2024re2llm} selects demonstrations for recommendation from a pool via PPO.
(9) A-LLMRec~\cite{kim2024large} enables an LLM to leverage collaborative knowledge through multi-stage training.
(10) TALLRec~\cite{bao2023tallrec} fine-tunes by providing a text description and selecting one target through LLM.
(11) LLaRA~\cite{liao2024llara} combines behavioral patterns with world knowledge for sequential recommendations.
(12) MoRE~\cite{qin2024enhancing} provides suitable reflections for recommendations in a multi-perspective and adaptive manner, decoupling and exploring users' explicit and implicit preferences while integrating collaborative information.

For a fair comparison, all baselines use the same backbone, the same processed dataset, and the same item candidate set. Note that TALLRec, A-LLMRec, and LLaRA specialize in \textbf{top-1} prediction because they \textbf{do not have ranking capabilities}. \\
\textbf{Evaluation Protocols.}
To evaluate ranking and recall capabilities, methods other than those specialized in top-1 prediction are required to output top-10 ranked lists, with ``nDCG@$k$'' and ``Recall@$k$'' used as metrics ($k=10$). To evaluate top-1 prediction performance, we adopt ``nDCG@$1$''. \\
\textbf{Implementation Details.}
Following~\cite{zhou2022filter, qin2024enhancing}, we randomly sample 1,000 users, with items remaining
unchanged for evaluation, and then split the training, validation, and test data based on the leave-one-out strategy for all 3 datasets respectively. Following~\cite{kim2024large,wang2024re2llm,qin2024enhancing}, we construct the candidate set by randomly sampling items with one ground-truth item, with the candidate set size being 50, 49 for negative items and 1 for ground-truth item. 

Following~\cite{zheng2024adapting,qin2024enhancing}, we implement traditional methods with RecBole \cite{zhao2021recbole} and adopt the Adam optimizer and grid search where the embedding size is tuned among \{32, 64, 128\}, the learning rate is among \{$1e^{-2}, 1e^{-3},1e^{-4}$\}, and the batch size is among \{1024, 2048, 4096\}.

Following~\cite{qin2024enhancing}, for all LLM-based methods, we adopt ``LLaMa3''\footnote{LLaMa-3-8B-Instruct.} \cite{dubey2024llama} as the default backbone and additionally involve LLMs\footnote{They are Phi-3-mini-4k-Instruct, Qwen2-7B-Instruct, Qwen2.5-7B-Instruct, and Mistral-7B-Instruct-v0.1, respectively.} such as ``Phi3''~\cite{abdin2024phi}, ``Qwen2''~\cite{yang2024qwen2technicalreport}, ``Qwen2.5''~\cite{yang2024qwen2}, and ``Mistral''~\cite{jiang2023mistral} for model-agnostic validation. 
We conducted user sampling identical to that of~\citep{qin2024enhancing} and made proper adjustments\footnote{Following~\cite{liao2024llara,kim2024large,qin2024enhancing}, we adjust TALLRec to enable prediction of the next items. We also construct candidate constraints for LC-Rec to avoid non-candidate outputs.
} to some of the baselines, in line with~\citep{qin2024enhancing}, to adapt to our scenarios. All experiments are run on a A800-PCIE-80GB GPU. We conduct 10 repeated trials for each experiment. Since LRGD only use ICL method to make recommendations, which means that training time comparison is not applicable for LRGD, so we do not make training efficiency experiment in the following parts.

\subsection{LRGD Equivalence Validation Experiment}

\begin{figure}[!t]
    % \centering
    \setlength{\belowcaptionskip}{0pt}
    \setlength{\abovecaptionskip}{0pt}
    \begin{subfigure}[h]{0.235\textwidth}
        \setlength{\belowcaptionskip}{3pt}
        \setlength{\abovecaptionskip}{0pt}
        \includegraphics[width=\textwidth]{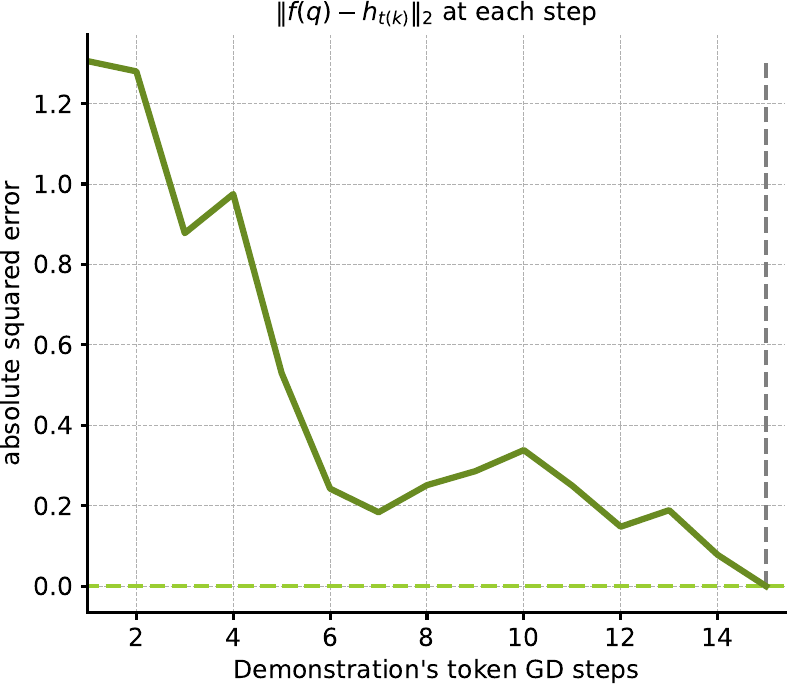}
        \caption{$N_D=15$, hit at $1^{st}$ output token}
        \label{fig:theo:1st}
    \end{subfigure}
    \hfill
    \begin{subfigure}[h]{0.235\textwidth}
        \setlength{\belowcaptionskip}{3pt}
        \setlength{\abovecaptionskip}{0pt}
        \includegraphics[width=\textwidth]{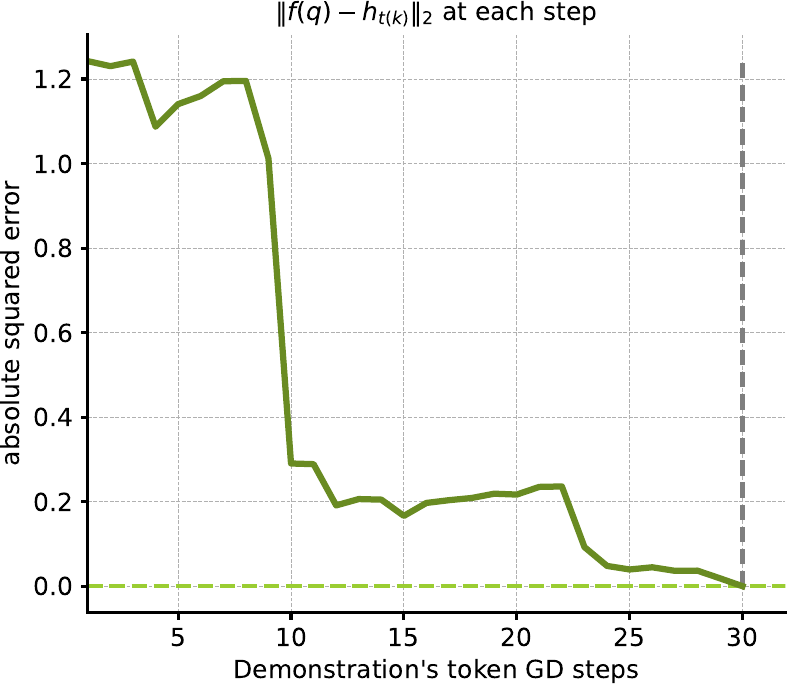}
        \caption{\footnotesize{$N_D=10$, hit at $3^{rd}$ output token}}
        \label{fig:theo:3rd}
    \end{subfigure}
    \caption{Equivalence validation experiment result of LRGD. Decoder-only model is set to be single-layer masked softmax attention with $d_i=11, d_o=1,\bm N_T=15, k=2$. The vertical axis in the figure represents the absolute squared error (SE) difference, while the horizontal axis shows the number of gradient descent steps executed for each demonstration token in $\bm X_D$.}
    \label{fig:theo}
\end{figure}

We design the following validation experiment to verify the equivalence of LRGD discussed in §\ref{sec:method:theory:understand} and the effectiveness of \( \mathrm{Effect}_D \) in §\ref{sec:method:theory:effective}, answering \textbf{RQ1}. We observe the difference between the real decoder-only model output \( \bm h_{t(k)} \) and the dual model output \( f(\bm q) \), derived from the decoder-only model's parameters. In Fig.~\ref{fig:theo:1st}, with 15 tokens in a good demonstration, \( f(\bm q) \) matches \( \bm h_{t(k)} \) on the first output token after one whole gradient descent step for each token. In contrast, in Fig.~\ref{fig:theo:3rd}, with 10 tokens in a bad demonstration, \( f(\bm q) \) only matches \( \bm h_{t(k)} \) on the third output token after three whole gradient descent steps for each token. Fig.~\ref{fig:theo} effectively demonstrates the equivalence of LRGD and the effectiveness of \( \mathrm{Effect}_D \).

\label{sec:exp:theo}

\subsection{Overall Recommendation Performance}

\begin{table}[tbp!]
\centering
\small
\caption{Comparison of top-1 performance between LRGD and the LLM baselines specialized in top-1  predictions. 
% For fairness, all methods use the same backbone, processed data, and candidate set. 
The best results are shown in bold font. 
% ``Imp'' indicates the percentage improvement of LRGD over the best baseline performances. 
% ``*'' denotes that the performance improvement of LRGD is statistically significant (\(p \leq 0.05\)).
LRGD significantly outperforms the LLM baseline specialized in top-1 predictions on the metric of top-1 prediction, demonstrating its superior performance.}
\label{tab:exp:top1}
\begin{tabular}{cccccc}
\hline
Dataset     & TALLRec & LLaRA & A-LLMRec  & LRGD            & Imp.     \\ \hline
Games       & 0.2031  & 0.2313 & 0.2127 & \textbf{0.2977} & 28.71\%* \\
Arts        & 0.2487  & 0.2799 & 0.2505 & \textbf{0.3005} & 7.36\%*  \\
Instruments & 0.1774  & 0.2110 & 0.1954 & \textbf{0.2496} & 18.29\%* \\ \hline
\end{tabular}
\end{table}

To answer \textbf{RQ2}, we validate the recommendation performance of the LRGD application,  separately comparing its ranking capability and top-1 prediction capability with baselines. 

For the evaluation of ranking capability, we analyze the ranking and recall metrics of LRGD against existing methods on three datasets, as shown in Tab.~\ref{tab:exp:overall}. Our findings show that LRGD significantly outperforms other LLM-based methods as well as all traditional approaches.

For the evaluation of top-1 prediction capability, we examine nDCG@$1$ on the same three datasets, as presented in Tab.~\ref{tab:exp:top1}. Even baselines specifically designed for top-1 prediction are outperformed by LRGD.

These comparisons highlight the superior effectiveness and practicality of LRGD.

\subsection{Model-Agnostic Validation}
\begin{table}[tbp!]
\small
\setlength{\tabcolsep}{2.6pt}
\caption{Model-agnostic validation on Games. %``ZSL'' refers to LLM-ICL recommendations made without demonstrations. 
% ``ICL'' uses demonstrations without optimization, and 
``LRGD'' applies demonstrations derived from two-stage iterative optimization. ``Imp'' indicates the percentage improvement of LRGD over ZSL performances. 
% LRGD significantly boosts recommendation performance across all LLMs, showcasing its model-agnostic benefits. 
% ``*'' denotes that the performance improvement of LRGD is statistically significant (\(p \leq 0.05\)).
}
\label{tab:exp:model-agnostic}
\begin{tabular}{@{}lllllll@{}}
\hline
\multicolumn{2}{l}{\textbf{Decoder-Only LLMs}}             & \multicolumn{1}{c}{LLaMa3} & \multicolumn{1}{l}{Phi3} & \multicolumn{1}{l}{Qwen2} & \multicolumn{1}{l}{Qwen2.5} & \multicolumn{1}{l}{Mistral} \\ \hline
\multirow{4}{*}{nDCG@1}  & ZSL       & 0.2318                     & 0.1049                   & 0.1918                    & 0.1938                      & 0.0679                      \\
                         & ICL       & 0.2337                     & 0.1139                   & 0.2038                    & 0.2418                      & 0.0739                      \\
                         & LRGD & \textbf{0.2977  }                   & \textbf{0.1359}                   & \textbf{0.2657}                    & \textbf{0.3716}                      & \textbf{0.1129}                      \\
                         & Imp.      & 28.43\%*                    & 29.55\%*                  & 38.53\%*                  & 91.74\%*                     & 66.27\%*                     \\ \hline
\multirow{4}{*}{nDCG@10} & ZSL       & 0.3602                     & 0.1896                   & 0.2639                    & 0.3226                      & 0.1558                      \\
                         & ICL       & 0.3761                     & 0.1968                   & 0.2904                    & 0.3685                      & 0.1608                      \\
                         & LRGD & \textbf{0.4306}                     & \textbf{0.2091}                   & \textbf{0.3733}                    & \textbf{0.4920}                       & \textbf{0.1893}                      \\
                         & Imp.      & 19.54\%*                    & 10.28\%*                  & 41.46\%*                   & 52.51\%*                     & 21.50\%*                     \\ \hline
\end{tabular}
\end{table}
To answer \textbf{RQ3}, we involve different LLM backbones. Tab.~\ref{tab:exp:model-agnostic} illustrates the enhancement in recommendation performance of LRGD with decoder-only LLMs. The degree of improvement varies across different LLMs, with a particularly significant boost observed in ``Qwen2.5''. These results underscore the flexibility and generalization ability of LRGD in improving recommendations across various decoder-only LLM architectures.

% \begin{figure}[t!] % 
%     \centering 
%     \setlength{\belowcaptionskip}{0pt}
%     \setlength{\abovecaptionskip}{0pt}
%     \includegraphics[width=0.45\textwidth]{fig/demo_similarity_perturbation_v4.pdf}
%     \caption{Comparison between before and after alleviating demonstration collapse via $\bm X_{D_{per}}$ in LRGD. The solid line represents the performance with the addition of \( X_{D_{per}} \) in the iteration. The dashed line represents the performance without the addition of \( X_{D_{per}} \). The left vertical axis shows the semantic similarity between the current and previous round's \( X_D \), while the right vertical axis represents the nDCG@10. The horizontal axis indicates the number of iterations. Error bar indicates the standard error.} 
%  \label{fig:exp:collapse}
% \end{figure}

\subsection{Abation study}
\begin{table}[tbp!]
\centering
\small
\setlength{\tabcolsep}{4pt}
\caption{Ablation comparison on Arts. LRGD adopts ``3-PDO''. ``1-PDO'' is equivalent to ``w/o $\bm X_{D_{\text{per}}}$''. ``$\mathrm{Inv}({X_T, X_D})$'' denotes the swapping of positions between $X_T$ and $X_D$, used to verify the necessity of rotation matrix $\bm R_i$ in LRGD.}
\label{tab:exp:ablation}
\begin{tabular}{lccccc}
\hline
\multicolumn{1}{l}{Metric} & \multicolumn{1}{l}{LRGD} & \multicolumn{1}{l}{2-PDO} & \multicolumn{1}{l}{1-PDO} & \multicolumn{1}{l}{w/o $X_{T_{reg}}$} & \multicolumn{1}{l}{$\mathrm{Inv}({X_T, X_D})$} \\ \hline
nDCG@1                     & \textbf{0.3005}          & 0.2937                          & 0.2907                      & 0.2839                                & 0.2644                                             \\
% nDCG@5                     & \textbf{0.3756}          & 0.3616                          & 0.3583                      & 0.3560                                 & 0.3502                                             \\
nDCG@10                    & \textbf{0.3962}          & 0.3804                          & 0.3782                      & 0.3760                                 & 0.3701                                               \\
% Recall@5                   & \textbf{0.4420}           & 0.4254                          & 0.4244                      & 0.4234                                & 0.4263                                             \\
Recall@10                  & \textbf{0.5063}          & 0.5015                          & 0.4907                      & 0.4868                                & 0.4927                                             \\ \hline
\end{tabular}
\end{table}
To answer \textbf{RQ4}, we conduct an ablation study to analyze the contribution of each component in our proposed model. The results are summarized in Tab.~\ref{tab:exp:ablation}, where we evaluate the performance of different variants of our model on Arts. 

It can be observed that:
(1) Reducing the paths of demonstration optimization (from 3-PDO to2-PDO) leads to performance drops, emphasizing the importance of $m$-path demonstration optimization in the iterative optimization process (§\ref{sec:method:app}).
(2) The absence of $X_{T_{\text{reg}}}$ and $X_{D_{\text{per}}}$ (1-PDO) further degrades performance, demonstrating their critical role in LRGD. 
(3)
% \textbf{Necessity of Rotation Matrix $\bm{R}_i$:} 
The $\mathrm{Inv}(X_T, X_D)$ variant, which swaps the positions of $X_T$ and $X_D$, shows the lowest performance among all metrics. This verifies the necessity of the consideration of positional encoding and rotation matrix $\bm{R}_i$ in LRGD.

In short, LRGD achieves the highest performance across all metrics, indicating that each component plays a crucial role.

\subsection{Analysis of Demonstration Collapse}
\begin{figure}[t!] % 
    \centering 
    \setlength{\belowcaptionskip}{-5pt}
    \setlength{\abovecaptionskip}{0pt}
    \includegraphics[width=0.43\textwidth]{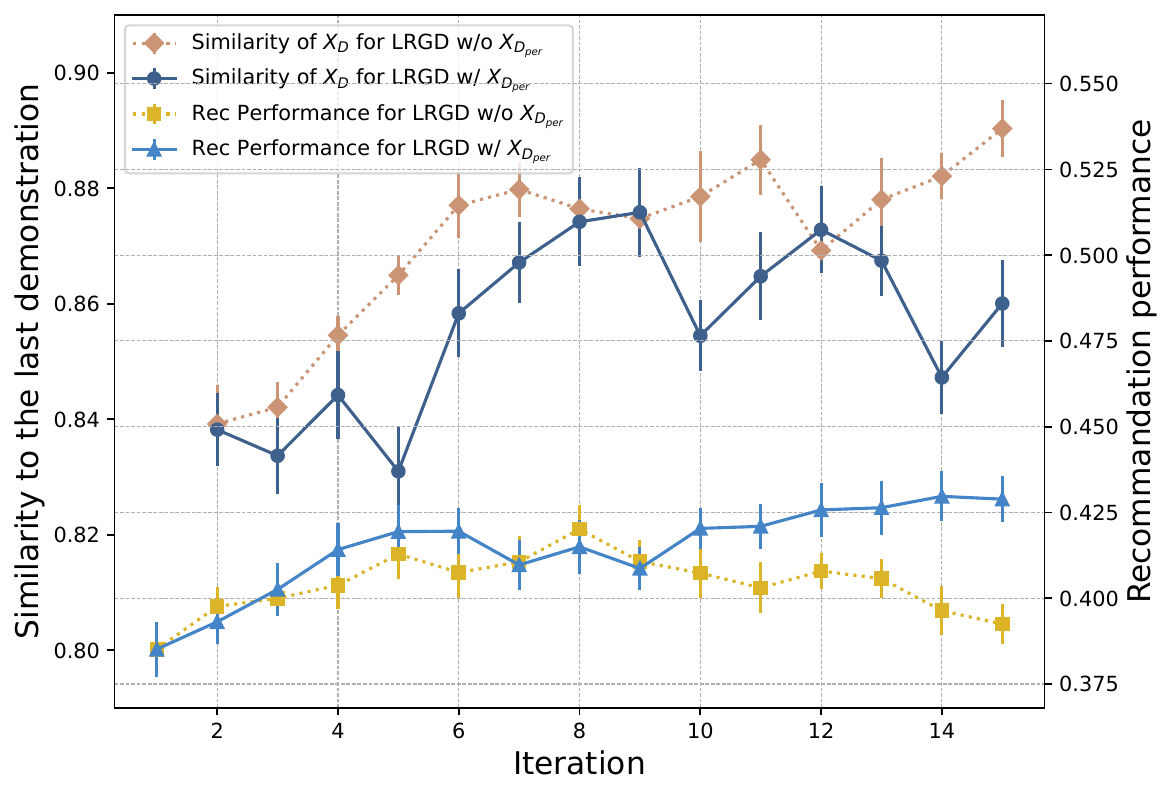}
    \caption{Comparison between before and after alleviating demonstration collapse via $\bm X_{D_{per}}$ in LRGD. The solid line represents the performance with \( X_{D_{per}} \) in the iteration. The dashed line represents the performance without \( X_{D_{per}} \). The left vertical axis shows the semantic similarity between the current and previous round's \( X_D \), while the right vertical axis represents the nDCG@10. The horizontal axis indicates the number of iterations. Error bar indicates the standard error.} 
 \label{fig:exp:collapse}
\end{figure}
To answer \textbf{RQ5}, we analyze the effectiveness of LRGD in mitigating the demonstration collapse issue (§\ref{sec:method:app}). Fig.~\ref{fig:exp:collapse} shows a comparison of performance and demonstration similarity before and after incorporating the $\bm{X}_{D_{per}}$ into LRGD. %The solid line and dashed line respectively represent iterative optimization process with perturbation $\bm{X}_{D_{per}}$ and without $\bm{X}_{D_{per}}$. 

It is evident that adding $\bm{X}_{D_{per}}$ significantly reduces the similarity between consecutive rounds, demonstrating its ability to introduce meaningful diversity and mitigate collapse. Furthermore, the nDCG@10 values consistently improve with the inclusion of $\bm{X}_{D_{per}}$, highlighting its positive impact on recommendation quality. 

These results confirm that LRGD effectively alleviates demonstration collapse by maintaining diversity and improving overall model performance.

% \subsection{Case Study}

\section{Conclusion}

In this paper, we address key challenges in ICL-based recommender systems: limited theoretical understanding, inadequate evaluation metrics, and suboptimal demonstration optimization. Our proposed LRGD model connects recommendation token generation with gradient descent dynamics. We introduce:(1) A novel evaluation metric $\mathrm{Effect}_D$. (2)A two-stage optimization framework with perturbations and regularizations for robustness. Extensice experiments on three real-world datasets confirm LRGD's theoretical equivalence and state-of-the-art performance, demonstrating both effectiveness and model-agnostic adaptability. The framework bridges theoretical principles with practical LLM-driven recommender systems.

\newpage
\bibliographystyle{ACM-Reference-Format}
\bibliography{acmsigir}
\newpage
\appendix
\section{Appendix: Connection between Attention and Gradient Descent}
\label{app:conn}
By exploring the connections between gradient descent and linear attention mechanisms, as well as bridging softmax and linear attention through kernel transformations, we provide the groundwork for the proposed method.

\subsection{Gradient Descent and Linear Attention: A Duality}
Recent studies \cite{ren2024towards, dai2023why, zhang2023batch, yao2024enhancing} have established a connection between linear attention and gradient descent on linear layers, interpreting the gradient descent process as the dual equivalent of linear attention. Consider the simplest single-layer linear model, defined as follows:
$$
f(x) = \bm W_0 \bm x,
$$
where $\bm W_0 \in \mathbb{R}^{d_o \times d_i}$ represents the weight matrix of the linear layer, and $\bm x \in \mathbb{R}^{d_i}$ is the input vector. During training, given an input sequence $[\bm x_k]_{k=1}^{N} \in \mathbb{R}^{d_i\times N}$ and corresponding label sequence $[\bm y_k]_{k=1}^{N} \in \mathbb{R}^{d_o\times N}$, the model is trained by minimizing the loss function $[\mathcal{L}(\hat{\bm y}_k, \bm y_k)]_{k=1}^{N}$, where $[\hat{\bm y}_k]_{k=1}^{N} = [\bm W_0 \bm x_k]_{k=1}^{N} \in \mathbb{R}^{d_o\times N}$ are the predicted outputs. Based on the current losses, an error signal $[\bm e_k]_{k=1}^{N} = [-\beta \frac{\partial \mathcal{L}}{\partial \bm y_k}]_{k=1}^{N} \in \mathbb{R}^{d_o\times N}$ (with $\beta$ being the learning rate) is computed and propagated back through the network during the backpropagation process, resulting in updates to the model parameters from $\bm W_0$ to $\bm W'$. The update rule is given by:
$$
\bm W' = \bm W_0 + \Delta \bm W = \bm W_0 + \sum_{k=1}^{N} \bm e_k \otimes \bm x_k.
$$
where $\otimes$ denotes the outer product of vectors.

Next, we examine the content of linear attention (LA). Let $\bm V=[\bm v_k]_{k=1}^{N}, \bm K =[\bm k_k]_{k=1}^{N}  \in \mathbb{R}^{d_o \times N}$ represent the value and key matrices, respectively. Given a query vector $\bm q\in \mathbb{R}^{d_o}$, linear attention can be expressed as:
$$
\mathrm{LA}(\bm V, \bm K, \bm q) = \bm V \bm K^\top \bm q  = \left( \sum_{k=1}^{N} \bm v_k \otimes \bm k_k \right) \bm q.
$$
For a trained linear layer model, during the test phase, given an input
$\bm x'\in \mathbb{R}^{d_i}$ 
we can derive the equivalent form of gradient descent and linear attention from the above equations:
$$
f(\bm x') = \bm W' \bm x' = (\bm W_0 + \sum_{k=1}^{N} \bm e_k \otimes \bm x_k) \bm x' = \bm W_0 \bm x' + \mathrm{LA}(\bm E, \bm X, \bm x').
$$
This demonstrates that the gradient descent process in training a linear model can be interpreted as a linear attention mechanism, with the model parameters $\bm W'$ adjusted by the error terms and the resulting attention computation during inference.

\subsection{Connecting Softmax Attention to Linear Attention with Kernels}
\label{sec:connect}
Due to the nonlinear normalization differences between softmax attention in the transformer decoder layers of LLMs and linear attention, we follow \cite{ren2024towards} to approximate softmax attention using kernel methods, with the aim of transforming the softmax component into an equivalent linear attention form. 

We begin by expressing the softmax function as the product of an unnormalized exponential term and a normalization vector \(\bm c\):
$$
\mathrm{softmax}(\bm X_1^\top \bm X_2) = \exp(\bm X_1^\top \bm X_2) \bm c,
$$
where \(\bm X_1 \in \mathbb{R}^{d_o \times d_1}\), \(\bm X_2 \in \mathbb{R}^{d_o \times d_2}\), \(\exp(\cdot)\) denotes element-wise exponentiation, and \(\bm c = \mathrm{diag}(\bm 1_{d_2 \times d_1} \exp(\bm X_1^\top \bm X_2))^{-1} \in \mathbb{R}^{d_2}\) is the normalization vector.

% For individual vectors $\bm x, \bm y \in \mathbb{R}^{d_o}$, we can rewrite $\mathrm{exp}(\bm x, \bm y)$ as:
% $$
% \mathrm{exp}(\bm x, \bm y) = e^{\bm x^\top \bm y} = e^{||\bm x||^2_2 / 2} \cdot e^{-||\bm x - \bm y||^2_2 / 2} \cdot e^{||\bm y||^2_2 / 2},
% $$
% where $e^{-||\bm x - \bm y||^2_2 / 2}$ represents the common Radial Basis Function (RBF) kernel with \(\sigma^2 = 1\). This kernel can be approximated using various kernel approximation techniques. A common approach is to use the following:
% $$
% e^{-||\bm x - \bm y||^2_2 / 2} = \phi_0(\bm x)^\top \phi_0(\bm y),
% $$
% where 
% $$
% \phi_0(\bm x) = \frac{1}{\sqrt{D}} \left( \sin(\bm u_1^\top \bm x), \dots, \sin(\bm u_{D/2}^\top \bm x), \cos(\bm u_1^\top \bm x), \dots, \cos(\bm u_{D/2}^\top \bm x) \right)^\top,
% $$
% and \(D\) is a constant (typically around 100), with \(\bm u_i\) being random vectors drawn from \(\mathcal{N}(0, \bm I_{d_o})\). The mapping \(\phi_0(\bm x)\) transforms \(\bm x \in \mathbb{R}^{d_o}\) to \(\mathbb{R}^{d_D}\).
% Thus, the exponential kernel approximation \(\phi(\bm x): \mathbb{R}^{d_o} \to \mathbb{R}^{d_D}\) can be written as:
% $$
% \mathrm{exp}(\bm x, \bm y) = \phi(\bm{x})^\top \phi(\bm{y})
% ,\ \phi(\bm{x}) = e^{||\bm x||^2_2 / 2} \phi_0(\bm x).
% $$
% This kernel approximation is consistent with the form proposed in \cite{choromanski2020rethinking} and will be utilized in this paper.
For individual vectors $\bm x, \bm y \in \mathbb{R}^{d_o}$, we can rewrite $\mathrm{exp}(\bm x, \bm y)$ with random Fourier mapping $\phi(\bm \cdot)$ based on the approximation of common RBF kernel\cite{rahimi2007random} with $\sigma^2=1$ as:
\begin{equation}
\label{app:eq:pre:kernel}
\mathrm{exp}(\bm x, \bm y) = e^{\bm x^\top \bm y} = \phi(\bm x)^\top \phi(\bm y),
\end{equation}
where 
$$
\phi(\bm x) = \frac{ e^{||\bm x||^2_2 / 2}}{\sqrt{D}} \left( \sin(\bm u_1^\top \bm x), .., \sin(\bm u_{D/2}^\top \bm x), \cos(\bm u_1^\top \bm x), .., \cos(\bm u_{D/2}^\top \bm x) \right)^\top,
$$
and \(D\) is a constant (typically around 100), with \(\bm u_i\) being random vectors drawn from \(\mathcal{N}(0, \sigma\bm I_{d_o})\). The mapping \(\phi(\bm x)\) transforms \(\bm x \in \mathbb{R}^{d_o}\) to \(\mathbb{R}^{d_D}\). $\phi(\bm \cdot)$ is consistent with the form proposed in \cite{choromanski2020rethinking} and will be utilized in this paper.

After applying the $\phi(\bm \cdot)$ to the individual vectors in \(\bm X_1\) and \(\bm X_2\) as described above, we obtain:
$$
\exp(\bm X_1^\top \bm X_2) = \phi(\bm X_1)^\top \phi(\bm X_2).
$$
Thus, the approximate form of the softmax function is:
$$
\mathrm{softmax}(\bm X_1^\top \bm X_2) = \phi(\bm X_1)^\top \phi(\bm X_2) \bm c,
$$
where \(\bm c = \mathrm{diag}(\bm 1_{d_2 \times d_1} \phi(\bm X_1)^\top \phi(\bm X_2))^{-1} \in \mathbb{R}^{d_2}\).

Finally, the output of the softmax attention can be rewritten as:
\begin{equation*}
% \label{eq:pre:kernel}
    \bm h = \bm V \, \mathrm{softmax}\left(\frac{\bm K^\top \bm q}{\sqrt{d_o}}\right) = c\bm V \, \phi(\bm K)^\top \phi(\bm q) = \mathrm{LA}(c \bm V, \phi(\bm K)^\top, \phi(\bm q)),
\end{equation*}
where
$c = \left( \bm 1_{N}^\top \phi(\bm K)^\top \phi(\bm q) \right)^{-1} \in \mathbb{R}^1
$, $\bm q \in \mathbb{R}^{d_i}$ is query vector.

This demonstrates how softmax attention in transformers can be approximated using kernel methods, transforming the softmax component into a linear attention form.

\newpage
\section{Appendix: Proof for Output of Attention and Its Dual Model}
\label{app:proof}
\textsc{Proof.~}\textit{To derive the gradient of \( \mathcal{L}_{ICL} \) with respect to \( \bm W \), we first compute the derivative:} 
\begin{equation}
\frac{\partial \mathcal{L}_{ICL}}{\partial \bm W} = -\frac{c \partial \sum_{i \in \mathcal{I}_D} (\bm v_i)^\top \bm W \phi(\bm k_i)}{\beta \partial \bm W} = -\frac{c}{\beta} \sum_{i \in \mathcal{I}_D} \bm v_i \phi(\bm k_i)^\top.
\end{equation}

\textit{Thus, the gradient \( \mathrm{grad}_{ICL} \) is equivalent to $\mathrm{grad}$:}
\begin{equation}
\mathrm{grad}_{ICL} = \beta \frac{\partial \mathcal{L}_{ICL}}{\partial \bm W} = -c \bm V_D \phi(\bm K_D)^\top = \mathrm{grad}.
\end{equation}

\textit{Consequently, the output of the dual model aligns with $\bm h_{t(k)}$:
}
\begin{equation}
\label{eq:equal}
\begin{aligned}
f(\bm q) &= \bm W \phi(\bm q) = {\bm W}_0 \phi(\bm q) - \mathrm{grad} \cdot \phi(\bm q) \\&= c \bm V_T \phi(\bm K_T)^\top \phi(\bm q) + c \bm V_D \phi(\bm K_D)^\top \phi(\bm q) = \bm h_{t(k)}.  
\end{aligned}
\end{equation}

This confirms that the generation process of $\bm h_{t(k)}$ in LRGD is equivalent to the dual model's gradient descent. 

\section{Appendix: Observation of LRGD Generalization for Multi-Layer Decoder-Only Language Models}
\label{app:obs:multi:layer}
To align with real-world applications of LLM-ICL Rec~\cite{bao2023tallrec,liao2024llara,qin2024enhancing}, we finally generalize LRGD to multi-layer decoder-only language models. Specifically, we can extend LRGD to the transformer decoder-only language model with  $L$  layers (GPT\cite{achiam2023gpt},  LLaMa\cite{dubey2024llama}, etc.). For the $k$-th newly output of the $L$-th layer $\hat{\bm x}_{t(k)}^{(L)}$, we have:
\begin{equation}
\begin{aligned}
&\hat{\bm x}_{t(k)}^{(L)} = \bm W_{\text{FFN}_1}^{(L)} \left[ \bm \Sigma_{\text{act}}^{(L)}  (\bm W_{\text{FFN}_2}^{(L)} \bm h_{t(k)}^{(L)} + \bm b_{\text{FFN}_2}^{(L)}) \right] + \bm b_{\text{FFN}_1}^{(L)} .\\
%&= \bm W_{\text{trm}}^{(l)} \bm V_T^{(l)} \phi(\bm K_T^{(l)})^\top \phi(\bm q^{(l)}) + \bm W_{\text{trm}}^{(l)} \bm V_D^{(l)} \phi(\bm K_D^{(l)})^\top \phi(\bm q^{(l)})\\ 
%&+\bm W_{\text{FFN}_1}^{(l)} \bm \Sigma_{\text{act}}^{(l)} \bm b_{\text{FFN}_2}^{(l)} + \bm b_{\text{FFN}_1}^{(l)}
%= \bm W_{\text{trm}}^{(l)'} \phi(\bm q) - \beta_{\text{trm}}^{(l)} \frac{\partial \mathcal{L}}{\partial \bm W_{\text{trm}}^{(l)}} \phi(\bm q^{(l)}) + \bm b_{\text{trm}}^{(l)}
\end{aligned}
\end{equation}
% Where:
% $$
% \bm V^{(l)} = [\bm W_v^{(l)} \bm x_i^{(l)}]_{i \in \mathcal{I}}, \ \bm K^{(l)} = [\bm R_i^{(l)} \bm W_k^{(l)} \bm x_i^{(l)}]_{i \in \mathcal{I}}
% $$
% $$
% \bm V_D^{(l)} = [\bm W_v^{(l)} \bm x_i^{(l)}]_{i \in \mathcal{I}_D}, \ \bm K_D^{(l)} = [\bm R_i^{(l)} \bm W_k^{(l)} \bm x_i^{(l)}]_{i \in \mathcal{I}_D}
% $$
% $$
% \bm q^{(l)} = \bm R_{N_T+N_D+k}^{(l)} \bm W_q^{(l)} \bm x^{(l)}_{N_T+N_D+k} ,\ \bm W_{\text{trm}}^{(l)} = c_{\text{trm}}^{(l)} \bm W_{\text{FFN}_1}^{(l)} \bm \Sigma_{\text{act}}^{(l)} \bm W_{\text{FFN}_2}^{(l)}
% $$
% $$
%  \quad \bm W_{\text{trm}}^{(l)'} = \bm W_{\text{trm}}^{(l)} \bm V_T^{(l)} \phi(\bm K_T^{(l)})^\top, \ \bm b_{\text{trm}}^{(l)} = \bm W_{\text{FFN}_1}^{(l)} \bm \Sigma_{\text{act}}^{(l)} \bm b_{\text{FFN}_2}^{(l)} + \bm b_{\text{FFN}_1}^{(l)}
% $$
% $$
% c_{\text{trm}}^{(l)} = \left( \bm 1_{N_T+N_D}^\top \phi(\bm K^{(l)})^\top \phi(\bm q^{(l)}) \right)^{-1},
% $$
The corresponding dual models for total $L$ layers are: 
\begin{equation}
\{f^{(l)}(\bm q^{(l)}) =\bm W^{(l)}_{\text{trm},0} \phi(\bm q^{(l)}) - \mathrm{grad}_{\text{trm}}^{(l)}\phi(\bm q^{(l)})  + \bm b_{\text{trm}}^{(l)}\}_{l=1}^{L}.
\end{equation}
 For $f^{(l)}(\bm q^{(l)})$, the constants are $,\bm b_{\text{trm}}^{(l)} =  \bm b_{\text{FFN}_1}^{(l)} + \bm W_{\text{FFN}_1}^{(l)} \bm \Sigma_{\text{act}}^{(L)} \bm b_{\text{FFN}_2}^{(l)} $, $\bm W_{\text{trm},0}^{(l)} = \hat{\bm W}_{\text{trm}}^{(l)} \bm V_T^{(l)} \phi(\bm K_T^{(l)})^\top$, where $\hat{\bm W}_{\text{trm}}^{(l)} = c_{\text{trm}}^{(l)} \bm W_{\text{FFN}_1}^{(l)} \bm \Sigma_{\text{act}}^{(l)} \bm W_{\text{FFN}_2}^{(l)}$. 
 
 We denote $\bm x_i^{(1)}$ as the original input from first layer. Each layer's input is connected with the output of last layer by connection matrix $\bm W_{\text{conn}}^{(l)}$, i.e. $\bm x^{(l)}_i = \bm W_{\text{conn}}^{(l)} \hat{\bm x}^{(l-1)}_i, \ (l > 1, i\in\mathcal{I})$.

The corresponding gradient $\mathrm{grad}_{\text{trm}}^{(l)}$ and  loss function $\mathcal{L}_{\text{trm}}^{(l)}$ for the  $l$ -th dual model $f^{(l)}(\bm q^{(l)})$ are:
\begin{equation}
\begin{aligned}
\mathrm{grad}_{\text{trm}}^{(l)} %= \beta_{\text{trm}}^{(l)} \frac{\partial \mathcal{L}}{\partial \bm W_{\text{trm}}^{(l)}} = -\bm W_{\text{trm}}^{(l)} \bm V_D^{(l)} \phi(\bm K_D^{(l)})^\top\\
& = -\sum_{i \in \mathcal{I}_D} \left( \bm W_{\text{trm}}^{(l)} \bm W_v^{(l)} \bm W_{\text{conn}}^{(l)} \hat{\bm x}^{(l-1)}_i \right) \otimes \phi \left( \bm R_i^{(l)} \bm W_k^{(l)} \bm W_{\text{conn}}^{(l)} \hat{\bm x}^{(l-1)}_i \right), \\
\mathcal{L}_{\text{trm}}^{(l)} & = -\frac{1}{\beta_{\text{trm}}^{(l)}} \left( \bm W_{\text{trm}}^{(l)}\bm V_D^{(l)} \right)^\top \left( \bm W^{(l)} \phi (\bm K_D^{(l)}) + \bm b^{(l)} \right).
\end{aligned}
\end{equation}
To obtain the final output $\hat{\bm x}_{t(k)}^{(L)}$, dual models need to perform gradient descent sequentially from layer 1 to $ L $. Only after all $ L $ layers sequentially complete individual steps can the entire process be considered as completing a full gradient descent for $\hat{\bm x}_{t(k)}^{(L)}$.
% \begin{figure}[!ht]
%     \centering
%     \includegraphics[width=0.45\textwidth]{fig/GQA_mechanism.pdf}
%     \caption{GQA mechanism}
%     \label{fig:GQA}
% \end{figure}

\section{Appendix: Observation of Generalization for Group Query Attention Scenarios}
\label{app:obs:GQA}
The latest decoder-only models commonly adopt the GQA mechanism, and the LRGD approach is still applicable. Specifically, the dual model of GQA follows a \textit{blockwise} gradient descent pattern: based on the definition of GQA, the number of query heads is not compressed and is assumed to be $ n \times g $, where $ n, g \in \mathbb{N}^+ $. The number of groups is set to $ g $, which is divisible by $ n \times g $, meaning the number of value and key heads is $ g $. The final attention output token $ \hat{\bm h}_{t(k)} \in \mathbb{R}^{d_o} $ is obtained by concatenating $ n \times g $ block outputs: $ [\bm h_{t(k)}^{(1)}, \bm h_{t(k)}^{(2)}, \dots, \bm h_{t(k)}^{(n \times g)}] $, where $ \bm h_{t(k)}^{(i)} \in \mathbb{R}^{d_o / (n \times g)} $, for all $ i \in [1, n \times g] $. The concatenation operation is:

$$
\hat{\bm h}_{t(k)} = \mathrm{Concat}(\bm W_{concat} [\bm h_{t(k)}^{(1)}, \bm h_{t(k)}^{(2)}, \dots, \bm h_{t(k)}^{(n \times g)}])
$$

where $ \bm W_{concat} \in \mathbb{R}^{(n \times g) \times \frac{d_o}{n \times g}} $ is the weight matrix for the concatenation of the heads, considered as a constant matrix. 

Then, for $ \forall i \in [0, n-1], \forall j \in [1, g] $, the $i \times g + j $-th sub-output token $ \hat{\bm h}_{t(k)}^{(s)} $ is given by: ($s = i \times g + j $)
\begin{equation*}
\begin{aligned}
&\hat{\bm h}_{t(k)}^{(s)} = \bm W_{concat}^{(s)} \bm h_{t(k)}^{(s)} = \bm W_{concat}^{(s)} \bm V^{(i)} \phi(\bm K^{(i)})^\top \phi(\bm q^{(s)}) 
%\\&= \bm W_{0}^{(s)} \phi(\bm q^{(s)}) - \beta^{(s)} \frac{\partial \mathcal{L}}{\partial \bm W^{(i)}} \phi(\bm q^{(s)})
\end{aligned}
\end{equation*}
% where:

% \begin{equation*}
% \begin{aligned}
% &\bm W_0^{(s)} = c^{(s)} \bm W_{concat}^{(s)} \bm V_T^{(i)} \phi(\bm K_T^{(i)})^\top,\\& c^{(s)} = \left( \bm 1_{N_T+N_D}^\top \phi(\bm K^{(i)})^\top \phi(\bm q^{(s)}) \right)^{-1}
% \end{aligned} 
% \end{equation*}

Thus, The corresponding $ n \times g $ \textit{blockwise} dual models are:$\{f^{(l)}(\bm q) = \bm W^{(l)} \phi(\bm q)=\bm W^{(l)}_{0}\phi(\bm q)-\mathrm{grad}^{(l)}\phi(\bm q)\}_{l=1}^{n \times g}$, and the final output is obtained after performing \textit{blockwise} gradient descent for each $ \hat{\bm h}_{t(k)}^{(s)} $. Constant part is $\bm W^{(s)}_0=c^{(s)} \bm W_{concat}^{(s)} \bm V_T^{(i)} \phi(\bm K_T^{(i)})^\top$. The gradient corresponding and loss function to the $ s $-th block output token is:

\begin{equation}
\mathrm{grad}^{(s)} =  - \sum_{m \in \mathcal{I}_D} \bm W_{concat}^{(s)} \bm v^{(i)}_m \otimes \phi(\bm k^{(i)}_m)^\top
\end{equation}

\begin{equation}
\mathcal{L}^{(s)} = -\frac{c^{(s)}}{\beta^{(s)}} \sum_{m \in \mathcal{I}_D} (\bm W_{concat}^{(s)} \bm v_m^{(i)})^\top \phi(\bm k_m^{(i)})
\end{equation}
As observed, in the GQA scenario, after completing the gradient descent for each block, the overall vector can be considered as completing one total gradient descent.

\end{document}